%% file: root.tex
\documentclass[final,5p,times,twocolumn,numbers]{elsarticle}
\input{preamble.tex}

\usepackage{tabularx}
\usepackage{algorithm}
\usepackage{algpseudocode}

\usepackage[subtle]{savetrees}

\usepackage{enumitem}
\setlist[itemize]{nosep,leftmargin=*}
\setlist[enumerate]{nosep,leftmargin=*}

\journal{Automatica}

\begin{document}
	
	\newdimen\origiwspc
	\newdimen\origiwstr
	\origiwspc=\fontdimen2=\fontdimen3{frontmatter}
		
		\title{\LARGE Power System CBFs}
		
		\author[inst1]{Abdallah Alalem B. Albustami}
		\affiliation[inst1]{organization={Vanderbilt University},
			addressline={Civil and Environmental Engineering Department},
			city={Nashville},
			postcode={37235},
			state={TN},
			country={USA}}
		\ead{abdallah.b.alalem.albustami@vanderbilt.edu}
		
		\author[inst1]{Ahmad F. Taha}
		\ead{ahmad.taha@vanderbilt.edu}
		
%		\cortext[1]{Corresponding author}

		\author[inst2]{Taylor T. Johnson}
		\affiliation[inst2]{organization={Vanderbilt University},
			addressline={Computer Science},
			city={Nashville},
			postcode={37235},
			state={TN},
			country={USA}}
		\ead{taylor.johnson@vanderbilt.edu}

		\begin{abstract}
			Control barrier functions (CBFs) have become a standard tool in safety critical-control systems. CBFs convert state constraints into real time control conditions that certify forward invariance (meaning that once the system starts in a safe region, it remains there for all future times) and minimally modify a nominal controller only when safety is at risk. In power systems, CBF based methods have been proposed for frequency and voltage safety, but they largely remain disconnected from three key features that are central to power system operation: differential algebraic equation (DAE) models that capture network power flow constraints, safety specifications involving algebraic variables such as bus voltages, and formal verification of the resulting closed loop system. This paper closes this gap by developing a CBF framework for power system DAE models that supports safety constraints on both dynamic and algebraic variables. The framework provides real time safety filtering through an optimization layer that wraps around an existing controller and minimally modifies its command to enforce safety. In addition, it provides formal verification (i.e., a mathematical guarantee that all admissible trajectories satisfy the prescribed safety constraints) through an offline reachability based certificate of safe operation. The result is a unified filter and verify methodology for enforcing and certifying frequency and voltage safety in power systems while preserving the DAE structure of the underlying model.
		\end{abstract}
		
		
	\end{frontmatter}

\section{Introduction and Paper Contributions}
\lettrine[lraise=0.1, nindent=0em, slope=-.5em]{P}{OWER} system frequency and voltage must remain within strict operational bands at all times. Frequency excursions beyond $\pm$0.5~Hz or bus voltage deviations outside the 0.95--1.05~p.u.\ range can trigger protection relay operations, cascade into widespread equipment disconnection, and in severe cases lead to system-wide blackouts~\cite{kundur1994power,ferc2025reliability}. These are not performance preferences but hard safety specifications that must hold under all credible operating conditions, including load disturbances, generation trips, and renewable fluctuations.

Power system controllers today are tuned to satisfy these constraints under typical conditions, but their correctness is validated almost exclusively through time domain simulation. An operator selects a handful of contingency scenarios, simulates the closed-loop response, and inspects whether the frequency and voltage trajectories stay within bounds~\cite{kundur1994power,sauer2017power}. This practice has two shortcomings. It covers only the initial conditions and disturbance profiles that happen to be tested, so it cannot certify safety across the continuum of operating states and disturbance realizations. And as renewable penetration grows and system inertia falls~\cite{milano2018foundations,ulbig2014impact}, the margin between normal operation and constraint violation shrinks, which makes validation by simulation increasingly inadequate.

Formal verification and safety-critical control offer complementary remedies. Safety-critical control, based on control barrier functions (CBFs), provides real-time safety filters that minimally modify a nominal controller to enforce constraint satisfaction~\cite{ames2016control,ames2019control}. Reachability analysis computes guaranteed outer approximations of the set of all possible future states, certifying that the system cannot enter unsafe regions~\cite{althoff2021set}. Together they form a filter-and-verify pipeline. The CBF filter enforces safety online, while reachability analysis certifies the closed-loop system offline. This paper develops such a pipeline for power systems modeled as differential algebraic equations (DAEs), the natural and physics-preserving representation that retains both dynamic device models and network power flow constraints.

\vspace{0.15cm}
\noindent \textbf{Related Literature.} The relevant literature spans four areas: \textit{(i)} safety-critical control based on barrier functions for power systems, \textit{(ii)} Lyapunov and set invariance methods for transient frequency and voltage safety, \textit{(iii)} formal verification and reachability analysis applied to power systems, and \textit{(iv)} CBF theory for constrained dynamical systems, including recent extensions to DAEs.

\vspace{0.15cm}
\noindent \textit{\textbf{1) Safety-Critical Control for Power Systems:}}
The application of barrier certificates and CBFs to power system control has gained significant traction in recent years. A reinforcement learning (RL) framework with barrier certificates for transient stability is proposed in~\cite{zhao2023barrier}, where a model-free RL controller discovers control actions that are subsequently filtered through a barrier certificate based on neural networks to satisfy frequency and voltage constraints. While effective, this approach relies on learned barrier functions whose validity is checked empirically rather than certified formally. An RL approach based on Lyapunov theory for primary frequency control is developed in~\cite{cui2022reinforcement}, which guarantees stability by construction and demonstrates strong performance on the IEEE~39-bus system. This line of work is extended in~\cite{yuan2024reinforcement}, where Lyapunov stability theory is combined with safety-critical control to derive sufficient conditions on distributed controller designs that ensure both stability and transient frequency safety, with a dynamic budget assignment that reduces conservatism relative to earlier work. Computationally efficient safe RL algorithms for voltage control are developed in~\cite{tabas2022computationally,shi2022stability}, formulating Volt-VAR problems as constrained Markov decision processes. Two patterns emerge across this body of work. The barrier or Lyapunov certificates are either learned from data, which introduces approximation gaps, or derived for reduced-order ODE models that eliminate the algebraic power flow constraints. Neither approach provides formal guarantees for the full DAE model, and none explicitly accounts for the model dependent relative degree that arises when the supervisory input acts through a cascade of controller dynamics.

\vspace{0.15cm}
\noindent \textit{\textbf{2) Lyapunov and Set Invariance Methods for Frequency and Voltage Safety:}}
Transient frequency safety, that is, keeping each bus frequency within a prescribed band, has been formalized using set invariance theory. In~\cite{zhangdistributed}, a distributed, Lipschitz continuous controller is synthesized that renders the safe frequency region forward invariant while simultaneously ensuring asymptotic stability, with the key insight that if stability and set invariance are both enforced, finite time convergence to the safe set follows automatically. This is extended in~\cite{zhang2020distributed} to bilayered architectures that decouple transient safety from steady state optimality. A neural Lyapunov control framework that jointly learns a Lyapunov function and control law via deep neural networks is presented in~\cite{zhao2021neural}, using a falsification module to validate candidate certificates against the nonlinear swing equation, but again on Kron-reduced ODE models without algebraic constraints. The voltage side is less developed. Voltage is an algebraic variable in DAE models, and its forward invariance cannot be addressed directly by standard Lyapunov methods designed for differential states. This structural gap motivates the DAE-aware CBF formulation in this work.

\vspace{0.15cm}
\noindent \textit{\textbf{3) Formal Verification and Reachability for Power Systems:}}
Reachability analysis computes guaranteed enclosures of all possible trajectories and has been applied to power system transient stability verification. The foundational algorithm for reachability analysis of nonlinear, semi-explicit, index-1 DAEs is developed in~\cite{althoff2013TAC} using conservative linearization and set propagation with zonotopes, with application to the IEEE~14-bus system. This is extended in~\cite{althoff2014TPWRS} to formal and compositional analysis of power systems, demonstrating that the IEEE~30-bus system can be verified for transient stability, variable renewable production, and bus voltage bounds without model simplification. Compositional techniques that decompose the power system into subsystems with interface bounds are developed in~\cite{el2017compositional} to further improve scalability. Standardized benchmarks for formal verification of power systems in the ARCH competition format are provided in~\cite{althoff2022benchmarks}, covering transient stability, region of attraction, and bus voltage verification for multiple IEEE systems. These works establish that reachability analysis is technically feasible for DAE power system models of moderate size. However, all of them verify \textit{open-loop} or passively controlled systems. None incorporates a real-time safety filter in the closed loop and verifies the resulting controlled system.

\vspace{0.15cm}
\noindent \textit{\textbf{4) CBF Theory and Extensions to DAEs:}}
The CBF framework for safety-critical control of ODE systems is now well established. CBF quadratic programs (CBF-QPs) for control-affine systems are formalized in~\cite{ames2016control}, yielding a computationally efficient QP solved at each time step. Subsequent work addresses higher relative degree systems. Exponential CBFs are introduced in~\cite{nguyen2016exponential}, HOCBFs of arbitrary order built from nested class-$\mathcal{K}$ recursions are developed in~\cite{xiao2021high}, a singularity-free HOCBF formulation that yields a locally Lipschitz control law and relaxes the uniform non-vanishing condition is given in~\cite{tan2021high}, and the multi-input case in which the relative degree of a barrier depends on the component of the control vector is analyzed in~\cite{xiao2022control}. Closed form solutions that avoid online QP solvers for single constraint, relative degree one CBF problems are given in~\cite{mestres2025explicit}. The robustness of HOCBFs to disturbances is studied in~\cite{tan2021high}, where the forward-invariant set is shown to be asymptotically stable and a matched-disturbance robustification is proposed. That analysis is for ODE systems and certifies robustness qualitatively, through stability of the set, rather than through an explicit and computable safe set inflation. Very recently, DAE-aware CBFs that account for the differential algebraic structure through projected vector fields on the constraint manifold are introduced in~\cite{zhang2026verification}, with necessary and sufficient conditions for geometric correctness and feasibility, and sum-of-squares certificates for polynomial systems. That work is demonstrated on robotic systems, not on networked power system DAE models where the algebraic constraints encode power flow physics across the entire network and the dynamic part is an assembly of heterogeneous device models whose relative degree to an operator facing supervisory input depends on the chosen model stack. CBF safety filters based on SOS methods are applied to a three-phase AC/DC power converter in~\cite{schneeberger2024advanced}, which operates at the component level and does not address network level DAE models with bus power balance constraints.

\vspace{0.15cm}
\noindent \textbf{Key Research Gaps.} Three gaps motivate this work. First, existing safety-critical control methods for power systems operate on Kron-reduced ODE models that eliminate the algebraic states, yet voltage safety, a primary operational concern, is a constraint on the algebraic variables that vanish under Kron reduction. Safety filtering on DAEs preserves the full coupling between dynamic device states and network voltages and so enforces both frequency and voltage constraints directly. Second, none of the existing barrier function approaches for power systems provides formal verification of the closed-loop safety guarantee on the full DAE. Third, the recent DAE-aware CBF theory~\cite{zhang2026verification} has not been applied to power system network models and does not treat a situation that arises routinely in practice: the operator facing supervisory input acts on the protected variable through a cascade of smooth controller dynamics, so the relative degree of each barrier is \textit{model dependent} and can exceed two. In some practical controller blocks the supervisory input enters the algebraic equations directly rather than through a pure integrator chain, which breaks the standard static QP CBF construction unless handled explicitly. A higher order DAE-CBF treatment that handles arbitrary relative degree uniformly on networked power system DAEs, stays sound when the input enters the algebraic part, and closes cleanly with offline reachability analysis is currently missing. Existing reachability engines applied to power systems, most notably CORA~\cite{Althoff2015ARCH}, have been demonstrated on open-loop or passively controlled DAEs up to the IEEE-30 scale; extending reachability to closed-loop DAE models that include a real-time safety filter based on CBFs has not been reported.

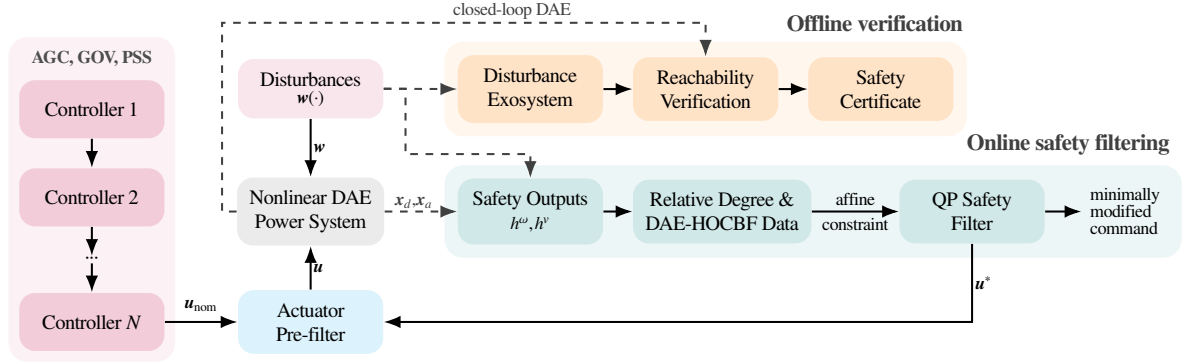
\begin{figure*}[t]
\centering
\resizebox{0.85\textwidth}{!}{
\begin{tikzpicture}[
    >=Latex,
    font=\footnotesize,
    node distance=7mm and 8mm,
    line/.style={-Latex, thick, black},
    dashline/.style={-Latex, thick, dashed, black!75},
    softbox/.style={
        rectangle,
        rounded corners=2.2mm,
        draw=none,
        align=center,
        inner sep=5pt
    },
    ctrl/.style={softbox, fill=purple!20, minimum width=20mm, minimum height = 8mm},
    pre/.style={softbox, fill=cyan!12, minimum width=20mm},
    plant/.style={softbox, fill=black!8, minimum width=20mm},
    online/.style={softbox, fill=teal!18, minimum width=20mm},
    offline/.style={softbox, fill=orange!22, minimum width=20mm},
    io/.style={font=\small},
    smalllab/.style={font=\scriptsize, inner sep=1pt},
    lane/.style={font=\small\bfseries, text=black!75},
    tag/.style={
        circle,
        draw=none,
        fill=black!12,
        minimum size=4.8mm,
        inner sep=0pt,
        font=\scriptsize\bfseries
    }
]
\node[ctrl] (c1) {Controller 1};
\node[ctrl, below=4mm of c1] (c2) {Controller 2};
\node[draw=none, below=4mm of c2, inner sep=0pt] (cdots) {$\cdots$};
\node[ctrl, below=4mm of cdots] (cN) {Controller $N$};
\node[smalllab, above=1.5mm of c1, text=black!70] (stacklbl) {\textbf{AGC, GOV, PSS}};
\node[pre, right=10mm of cN, minimum width=20mm] (pref) {Actuator\\Pre-filter};
\node[plant, above=6mm of pref] (dae) {Nonlinear DAE\\Power System};

\draw[line] (c1.south) -- (c2.north);
\draw[line] (c2.south) -- (cdots.north);
\draw[line] (cdots.south) -- (cN.north);
\draw[line] (cN.east) -- node[smalllab, above=1mm] {$\m{u}_{\mathrm{nom}}$} (pref.west);
\draw[line] (pref.north) -- node[smalllab, right] {$\m{u}$} (dae.south);

\begin{scope}[on background layer]
\node[
    rounded corners=2.6mm,
    fill=purple!5,
    line width=1.4pt,
    inner sep=4pt,
    fit=(c1)(c2)(cdots)(cN)(stacklbl)
] (ctrlfit) {};
\end{scope}

% Disturbance block
\node[softbox, fill=purple!10, minimum width=20mm, above=8mm of dae] (dist) {Disturbances\\[-1mm]\scriptsize $\m{w}(\cdot)$};
\draw[line] (dist.south) -- node[smalllab, right] {$\m{w}$} (dae.north);

\node[online, right =10mm of dae, minimum width=20mm] (bar) {Safety Outputs\\ \scriptsize $h^\omega,\;h^v$};

\node[online, right=4mm of bar, minimum width=20mm] (hocbf) {Relative Degree \&\\DAE-HOCBF Data};

\node[online, right=12mm of hocbf, minimum width=20mm] (qp) {QP Safety\\Filter};

\node[draw=none, right=5mm of qp, align=left] (cmdstar) {\scriptsize minimally\\[-1mm]\scriptsize modified\\[-1mm]\scriptsize command};

\draw[dashline] (dae.east) -- node[smalllab, pos=0.4, above] {$\m{x}_d,\m{x}_a$} (bar.west);
\draw[dashline] (dist.east) -| ++ (3mm,-9mm) -| ++ (5mm,0mm) -| (bar.north);

\draw[line] (bar.east) -- (hocbf.west);
\draw[line] (hocbf.east) -- node[smalllab, above=0.4mm] {\scriptsize affine} node[smalllab, below=0.4mm] {\scriptsize constraint} (qp.west);
\draw[line] (qp.east) -- (cmdstar.west);

\draw[line] (qp.south) |- node[smalllab, pos=0.25, right] {$\m{u}^{\ast}$} (pref.east);

\begin{scope}[on background layer]
\node[
    rounded corners=2.6mm,
    fill=teal!7,
    line width=1.4pt,
    inner sep=5pt,
    fit=(bar)(hocbf)(qp)(cmdstar)
] (onlinefit) {};
\end{scope}

\node[lane,  above right=0mm and 20mm of onlinefit.north] {Online safety filtering};

\node[offline, above=7.5mm of bar, minimum width=20mm] (exo) {Disturbance\\Exosystem};

\node[offline, right=4mm of exo, minimum width=20mm] (reach) {Reachability\\Verification};

\node[offline, right=4mm of reach, minimum width=20mm] (cert) {Safety\\Certificate};

\draw[line] (exo.east) -- (reach.west);
\draw[line] (reach.east) -- (cert.west);

\draw[dashline] (dae.west) -| ++ (-2mm,26mm) -| node[smalllab, pos=0.30, above] {\scriptsize closed-loop DAE} (reach.north);

\draw[dashline] (dist.east) -- (exo.west);

\begin{scope}[on background layer]
\node[
    rounded corners=2.6mm,
    fill=orange!7,
    line width=1.4pt,
    inner sep=5pt,
    fit=(exo)(reach)(cert)
] (offlinefit) {};
\end{scope}

\node[lane, above right=0mm and 10mm of offlinefit.north] {Offline verification};
\end{tikzpicture}}
\caption{Overview of the proposed filter-and-verify framework. An existing stack of power system controllers generates the nominal command. When needed, an actuator pre-filter is included so that the implemented plant fits the DAE-HOCBF construction. The online layer evaluates safety outputs, constructs the DAE-HOCBF constraints using model dependent relative degree information, and solves a QP that minimally modifies the nominal command. Offline, the same implemented closed-loop architecture is combined with the disturbance exosystem and verified by reachability analysis.}
\label{fig:framework_overview} \vspace{-0.4cm}
\end{figure*}

\vspace{0.15cm}
\noindent \textbf{Paper Contributions.} The contributions are stated for the class of smooth semi-explicit index-1 power system DAEs in which, after an actuator pre-filter where necessary, the supervisory input enters only the differential part and the reduced dynamics on the constraint manifold are control-affine. The precise structural form needed for the theory is introduced at the start of Section~\ref{sec:CBF}.
\begin{itemize}[leftmargin=*]
\item A DAE-HOCBF framework that uses a manifold-projected gradient to represent derivatives of barrier functions along DAE trajectories.
\item An actuator pre-filter construction for models in which the supervisory input enters the algebraic equations. The pre-filter is a dynamic extension of the plant, is part of the implemented control architecture, and renders the augmented system amenable to the static QP DAE-HOCBF construction.
\item A real-time QP safety filter that wraps around any nominal controller, accommodates barriers of heterogeneous relative degrees, and remains an ordinary strictly convex QP. For box constrained inputs we give an exact support function feasibility test for each robust DAE-HOCBF row, bound the slack induced and sampled-data safety inflation, and prove local Lipschitz regularity of the general soft QP filter. The single-row, inactive-box case also admits a closed form soft QP update.
\item A reachability formulation in which the disturbance derivatives appearing in the high order CBF condition are encoded as an integrator chain exosystem, so the verified closed-loop object is an index-1 DAE with augmented state and bounded exogenous input. We implement a reachability engine based on zonotopes for this closed-loop DAE and demonstrate certified safety under forecast uncertainty disturbance classes on the Kundur two-area and IEEE 39-bus systems.
\item Case studies on standard power system benchmarks with exciter, governor, and stabilizer dynamics, showing how the relative degree of each safety output is computed for the chosen model stack and how the corresponding HOCBF order is instantiated in the filter.
\end{itemize}

\noindent \textbf{Paper Organization.} Section~\ref{sec:prelim} presents the generic power system DAE model, safety specifications, and problem formulation. Section~\ref{sec:CBF} develops the DAE-HOCBF framework, including the actuator pre-filter construction. Section~\ref{sec:CORA} describes the reachability verification procedure. Section~\ref{sec:case_studies} presents case studies, and Section~\ref{sec:conclusion} concludes.

\section{Preliminaries: DAE Model, Assumptions, and Problem Formulation} \label{sec:prelim}

\noindent\textbf{Paper Notation.} Real $n$-vectors and real $m\times n$ matrices are written $\mathbb{R}$, $\mathbb{R}^n$, $\mathbb{R}^{m\times n}$. Throughout, $\m{x}_d\in\mathbb{R}^{n_d}$ denotes the dynamic state, $\m{x}_a\in\mathbb{R}^{n_a}$ the algebraic state, $\m{u}\in\mathbb{R}^{n_u}$ the supervisory input, and $\m{w}\in\mathbb{R}^{n_w}$ the exogenous disturbance, with $\m{x}=[\m{x}_d^\top,\m{x}_a^\top]^\top\in\mathbb{R}^n$ and $n:=n_d+n_a$. Vector fields and constraint maps are typed with their codomains, e.g.\ $\m{f}_0:\mathbb{R}^{n+n_w}\to\mathbb{R}^{n_d}$, $\m{B}_d:\mathbb{R}^{n+n_w}\to\mathbb{R}^{n_d\times n_u}$, $\m{g}:\mathbb{R}^{n+n_u+n_w}\to\mathbb{R}^{n_a}$. Their Jacobians satisfy $\m{J}_a=\partial \m{g}/\partial \m{x}_a\in\mathbb{R}^{n_a\times n_a}$, $\m{J}_d=\partial \m{g}/\partial \m{x}_d\in\mathbb{R}^{n_a\times n_d}$, and $\partial \m{g}/\partial \m{w}\in\mathbb{R}^{n_a\times n_w}$. Each safety output is a scalar map $h_j:\mathbb{R}^n\to\mathbb{R}$, the manifold-projected gradient is $\m{p}^\top\in\mathbb{R}^{1\times n_d}$, and the disturbance history vector is $\m{w}^{[\ell]}:=(\m{w},\dot{\m{w}},\ldots,\m{w}^{(\ell)})\in\mathbb{R}^{(\ell+1)n_w}$. The HOCBF coefficient maps satisfy $A_r\!:\!\to\!\mathbb{R}$, $\m{B}_r\!:\!\to\!\mathbb{R}^{n_u}$, $\m{\Gamma}_r\!:\!\to\!\mathbb{R}^{n_w}$, $\pi_{r-1}\!:\!\to\!\mathbb{R}$, with domains specified at first use. Class-$\mathcal{K}$ functions are continuous strictly increasing scalar maps $\alpha:\mathbb{R}\to\mathbb{R}$ with $\alpha(0)=0$. The Euclidean norm is $\|\cdot\|$ with dual $\|\cdot\|_*$, and $L_{\m{X}}\phi=(\partial\phi/\partial\m{\zeta})\m{X}$ denotes the Lie derivative of $\phi$ along $\m{X}$.

This section introduces the power system DAE model used throughout, states the operating constraints of interest, and formulates the safety filtering and verification problem. The presentation is kept general so that both differential state and algebraic state safety specifications are explicit before the barrier construction is specialized in Section~\ref{sec:CBF}.

\subsection{Power System DAE Model}\label{subsec:generic_model}

We adopt the standard semi-explicit DAE representation of power systems~\cite{kundur1994power}, in which the dynamic and algebraic variables of all network, machine, and controller components are stacked into two vectors
$\m{x}_d \in \mathbb{R}^{n_d}, \m{x}_a \in \mathbb{R} ^{n_a}.$
These collect, respectively, the \textit{dynamic states} declared by each device (rotor angles, speeds, flux states, exciter and governor states, stabilizer states, inverter-control states) and the \textit{algebraic states} declared by each device together with the bus voltage magnitudes and angles (machine stator algebraics, controller block outputs, reference variables, and the network power balance variables). We denote by $\m{u} \in \mathbb{R}^{n_u}$ the supervisory input that the safety filter manipulates, typically a vector of operator facing reference signals, and by $\m{w} \in \mathbb{R}^{n_w}$ the exogenous power injections (loads and non-dispatchable generation) and any other uncontrolled quantities.

Between limiter and switching events, the global model takes the semi-explicit DAE form
\begin{equation}\label{eq:DAE}
    \dot{\m{x}}_d = \m{f}(\m{x}_d, \m{x}_a, \m{u}, \m{w}), \qquad \m{0} = \m{g}(\m{x}_d, \m{x}_a, \m{u}, \m{w}),
\end{equation}
where $\m{f}$ aggregates all dynamic equations and $\m{g}$ aggregates all algebraic equations, including stator relations, internal controller algebraics, block interconnection equations, and the network active and reactive power balance at every bus, with a slack-bus angle constraint appended to make the algebraic system square. With $\m{x} = [\m{x}_d^\top \; \m{x}_a^\top]^\top$,
\begin{equation}\label{eq:DAE_compact}
    \m{E}\dot{\m{x}} = \m{F}(\m{x},\m{u},\m{w}), \qquad \m{E} = \begin{bmatrix} \m{I}_{n_d} & \m{0} \\ \m{0} & \m{0} \end{bmatrix}.
\end{equation}
Formulation~\eqref{eq:DAE} allows the supervisory input to enter both the differential and algebraic parts, covering both common actuation patterns: references that drive dynamic states of a controller block, and references that feed directly into an algebraic block interconnection.

The next assumption isolates the smooth index-1 regime on which the subsequent safety filter and reachability arguments are built.

\begin{asmp}[Smoothness and Index-1 Structure]\label{assump:main}
The following conditions hold in a neighborhood $\mathcal{N}$ of a nominal operating point $\m{z}_0 = (\m{x}_{d,0}, \m{x}_{a,0}, \m{u}_0, \m{w}_0)$.
\begin{enumerate}[label=(\roman*)]
    \item The algebraic Jacobian $\m{J}_a := \partial \m{g}/\partial \m{x}_a$ is nonsingular on $\mathcal{N}$, so~\eqref{eq:DAE} is a semi-explicit index-1 DAE with a locally unique algebraic solution $\m{x}_a = \m{\varphi}(\m{x}_d, \m{u}, \m{w})$.
    \item The maps $\m{f}$ and $\m{g}$ are $C^s$-smooth on $\mathcal{N}$ with $s \geq r_{\max}+1$, where $r_{\max}$ is the largest relative degree considered in the barrier specifications below.
    \item The exogenous signal $\m{w}(\cdot)$ belongs to the class
    \begin{align}\label{eq:W_class}
    \mathcal{W}_{r_{\max}} &:= \bigl\{\m{w}:[0,\infty)\to\mathbb{R}^{n_w} \;\big|\; \m{w}^{(\ell)} \text{ abs.\ cont., } \ell=0,\ldots,\notag \\
    &\qquad r_{\max}-1,\m{w}^{(r_{\max})}\in L^\infty\bigr\},
    \end{align}
    with bounds $\|\m{w}^{(\ell)}(t)\| \leq \bar{w}_\ell$ for scalars $\bar{w}_\ell\geq 0$, $\ell=0,\ldots,r_{\max}$, and $\m{w}(t) \in \mathcal{W}\subset\mathbb{R}^{n_w}$ compact. For any integer $q\leq r_{\max}$, we write $\mathcal{W}_q$ for the analogous class with $r_{\max}$ replaced by $q$.
    \item The control inputs obey actuator limits $\m{u}_{\min} \leq \m{u}(t) \leq \m{u}_{\max}$ componentwise.
\end{enumerate}
\end{asmp}

Conditions \textit{(i)} and \textit{(ii)} are standard for power system DAE analysis and hold at feasible power flow solutions with well-conditioned network Jacobians~\cite{sauer2017power}. Condition \textit{(iii)} fixes the disturbance regularity needed when safety outputs are differentiated up to order $r_{\max}$ and assembled into the higher order barrier recursion. The first $r_{\max}-1$ derivatives are absolutely continuous, while the top derivative need only be essentially bounded. Over the transient horizons of interest this is consistent with the smooth evolution of load profiles and renewable injections. Condition \textit{(iv)} reflects saturation limits of exciters, turbine valves, and inverter currents.

Practical device libraries also include hard limiters, deadbands, anti-windup saturation, and discrete toggles, so the closed-loop model is generally piecewise-smooth or hybrid. Our analysis applies on smooth operating regions between switching surfaces. Extending the framework to piecewise-smooth and hybrid regimes is outside the scope of this paper.

\subsection{Safety Specifications}\label{subsec:safety}

With the model and regularity assumptions in place, we state the operating constraints that the safety filter must enforce.

\begin{mydef}[Power System Safety Specifications]\label{def:safety}
A trajectory $(\m{x}_d(t), \m{x}_a(t))$ of~\eqref{eq:DAE} is \textit{safe} if for all $t \geq 0$:
\begin{enumerate}[label=(\roman*)]
    \item \textbf{Frequency safety:} $|\omega_i(t) - \omega_0| \leq \Delta\omega_{\max}$ for every generator-side speed state $\omega_i \in \m{x}_d$,
    \item \textbf{Voltage safety:} $v_{\min} \leq v_k(t) \leq v_{\max}$ for every bus voltage magnitude $v_k \in \m{x}_a$,
\end{enumerate}
where $\Delta\omega_{\max}$ is the maximum allowable frequency deviation and $v_{\min}$, $v_{\max}$ are the voltage magnitude limits.
\end{mydef}

These constraints define the safe set
\begin{equation}\label{eq:safe_set}
    \mathcal{C} = \bigcap_{j=1}^{N_h} \bigl\{(\m{x}_d, \m{x}_a) \mid h_j(\m{x}_d, \m{x}_a) \geq 0\bigr\},
\end{equation}
with natural choices $h_{\omega,i}(\m{x}_d) = \Delta\omega_{\max}^2 - (\omega_i - \omega_0)^2$ for frequency and $h_{v,k}(\m{x}_a) = (v_k - v_{\min})(v_{\max} - v_k)$ for voltage. The frequency constraint is a function of a differential state and the voltage constraint a function of an algebraic state. This distinction, together with the model stack in use, determines the relative degree of each barrier and the structure of the safety filter.

\subsection{Problem Formulation}\label{subsec:problem}

These operating limits lead to two coupled tasks: enforcing safety online with minimal modification of a nominal controller, and certifying offline that the resulting closed-loop system stays safe over the disturbance class of interest.

\begin{problem}[Safety Filtering and Verification for DAE Power Systems]\label{prob:main}
Given~\eqref{eq:DAE} with specifications from Definition~\ref{def:safety}, a nominal controller $\m{u}_{\emph{nom}}$, actuator bounds, and an admissible disturbance class $\mathcal{W}_{r_{\max}}$:
\begin{enumerate}[label=(\roman*)]
    \item Design a real-time safety filter $\m{u}^*(t) = \mathcal{F}(\m{x}_d(t), \m{x}_a(t), \m{w}^{[r_{\max}-1]}(t), \m{u}_{\emph{nom}}(t))$ that keeps closed-loop DAE trajectories in $\mathcal{C}$ while minimizing $\|\m{u}^* - \m{u}_{\emph{nom}}\|$.
    \item Verify, for all algebraically consistent $\m{x}(0) \in \mathcal{X}_0 \subset \mathcal{C}$ and all $\m{w}(\cdot) \in \mathcal{W}_{r_{\max}}$, that the forward reachable set of the closed-loop DAE does not intersect $\mathbb{R}^n \setminus \mathcal{C}$ on $[0,T]$.
\end{enumerate}
\end{problem}

\noindent The generic model~\eqref{eq:DAE} lets the supervisory input appear in either the differential or the algebraic equations. The safety filter construction developed next is stated for implemented plants in which the input appears only in the differential part and the reduced dynamics on the constraint manifold are control-affine, which is exactly the setting in which the barrier constraints become static inequalities in the supervisory input. When a practical controller places the input inside an algebraic block, Section~\ref{subsec:prefilter} augments the plant with an actuator pre-filter and applies the same theory to the resulting implemented plant.

\section{DAE-Aware Control Barrier Functions and Safety Filter Design} \label{sec:CBF}

We now develop the DAE-aware HOCBF construction used for online safety filtering. We first state the structural form needed for the derivations, then show how models with algebraic input dependence are converted into that form, and finally derive the higher order barrier conditions and the QP filter.

\begin{mydef}[Standard DAE-HOCBF Form]\label{def:std_form}
A semi-explicit DAE is in \emph{standard DAE-HOCBF form} on a domain $\mathcal{N}_{\mathrm{std}}\subseteq\mathbb{R}^{n+n_w}$ if \begin{subequations}\label{eq:std_form}
\begin{align}
    \dot{\m{x}}_d &= \m{f}_0(\m{x}_d, \m{x}_a, \m{w}) + \m{B}_d(\m{x}_d, \m{x}_a, \m{w})\,\m{u}, \label{eq:std_form_a}\\
    \m{0} &= \m{g}(\m{x}_d, \m{x}_a, \m{w}), \label{eq:std_form_b}
\end{align}
\end{subequations}
with $\m{f}_0:\mathbb{R}^{n+n_w}\to\mathbb{R}^{n_d}$, $\m{B}_d:\mathbb{R}^{n+n_w}\to\mathbb{R}^{n_d\times n_u}$, $\m{g}:\mathbb{R}^{n+n_w}\to\mathbb{R}^{n_a}$ all $C^s$-smooth ($s \geq r_{\max}+1$), and $\partial \m{g}/\partial \m{x}_a\in\mathbb{R}^{n_a\times n_a}$ nonsingular on $\mathcal{N}_{\mathrm{std}}$.\end{mydef}

Form~\eqref{eq:std_form} has two distinguishing features. The supervisory input $\m{u}$ enters only the differential part ($\partial \m{g}/\partial \m{u} \equiv \m{0}$), which guarantees that derivatives of $\m{u}$ do not appear when barrier functions are differentiated along trajectories, so the HOCBF construction yields a static QP in $\m{u}$. The differential dynamics are control-affine, which guarantees that the resulting constraint is affine in $\m{u}$.

When the original plant~\eqref{eq:DAE} is not in standard form, most commonly because the supervisory input enters an algebraic controller block so that $\partial \m{g}/\partial \m{u} \not\equiv \m{0}$, we first augment the plant with an actuator pre-filter and apply the theory below to that implemented augmented plant.

\subsection{Actuator Pre-Filter for Models with Algebraic Input Dependence}\label{subsec:prefilter}

We begin with the case that prevents a direct static QP construction, namely algebraic dependence on the supervisory input.

Suppose the original plant~\eqref{eq:DAE} has $\partial \m{g}/\partial \m{u} \not\equiv \m{0}$. Differentiating $\m{g}(\m{x}_d(t),\m{x}_a(t),\m{u}(t),\m{w}(t))=\m{0}$ produces a $(\partial\m{g}/\partial \m{u})\dot{\m{u}}$ term, so $\dot{\m{x}}_a$ depends on $\dot{\m{u}}$, and this propagates into the chain rule expansion of any barrier depending on $\m{x}_a$. The CBF inequality then becomes a constraint in $(\m{u},\dot{\m{u}})$ rather than in $\m{u}$ alone.

We resolve this by dynamic extension. Let $\m{T}=\m{T}^\top\in\mathbb{R}^{n_u\times n_u}$ be a diagonal positive definite matrix of pre-filter time constants, chosen small enough to preserve the effective actuation bandwidth, and introduce a new supervisory command $\m{\nu} \in \mathbb{R}^{n_u}$ driving the first-order pre-filter
\begin{equation}\label{eq:prefilter}
    \m{T}\dot{\m{u}} = -\m{u} + \m{\nu}.
\end{equation}
Augment the dynamic state as $\tilde{\m{x}}_d := [\m{x}_d^\top, \m{u}^\top]^\top$, keep $\tilde{\m{x}}_a := \m{x}_a$, and obtain the augmented DAE
\begin{subequations}\label{eq:aug_DAE}
\begin{align}
    \dot{\m{x}}_d &= \m{f}(\m{x}_d, \m{x}_a, \m{u}, \m{w}), \label{eq:aug_DAE_a}\\
    \dot{\m{u}} &= \m{T}^{-1}(-\m{u} + \m{\nu}), \label{eq:aug_DAE_b}\\
    \m{0} &= \m{g}(\m{x}_d, \m{x}_a, \m{u}, \m{w}). \label{eq:aug_DAE_c}
\end{align}
\end{subequations}
The new input $\m{\nu}$ does not appear in~\eqref{eq:aug_DAE_c}, so $\partial \m{g}/\partial \m{\nu} \equiv \m{0}$, and the algebraic Jacobian $\partial \m{g}/\partial \tilde{\m{x}}_a = \m{J}_a$ inherits nonsingularity. The augmented differential dynamics are control-affine in $\m{\nu}$, which enters only through~\eqref{eq:aug_DAE_b} with coefficient $\m{T}^{-1}$, while $\m{f}(\m{x}_d,\m{x}_a,\m{u},\m{w})$ in~\eqref{eq:aug_DAE_a} is absorbed into the drift with $\m{u}$ now a state. Hence~\eqref{eq:aug_DAE} is in the standard form of Definition~\ref{def:std_form}.

\vspace{0.1cm}
\noindent\textbf{Pre-filter is a dynamic extension.} The pre-filter~\eqref{eq:prefilter} adds $n_u$ states, imposes first-order low-pass dynamics between $\m{\nu}$ and the physical input $\m{u}$, and alters both open-loop and closed-loop reachable sets relative to~\eqref{eq:DAE}. We treat~\eqref{eq:prefilter} as part of the implemented control architecture throughout: the safety filter computes $\m{\nu}$, the pre-filter produces $\m{u}$, and the plant receives $\m{u}$. All subsequent results, including the reachability certificate, concern the augmented closed-loop system~\eqref{eq:aug_DAE} rather than the original~\eqref{eq:DAE}.

\vspace{0.1cm}
\noindent\textbf{Relative degree bookkeeping.} In the augmented model $\m{u}$ is a dynamic state. For any barrier $h(\m{x}_d,\m{x}_a)$, the relative degree with respect to $\m{u}$, understood as the number of derivatives of $h$ along~\eqref{eq:aug_DAE} before $\m{u}$ appears while treating $\m{u}$ as a state, is well defined. The relative degree with respect to the new command $\m{\nu}$ is exactly one higher, since $\m{\nu}$ enters $\m{u}$ through the single integrator~\eqref{eq:aug_DAE_b}. When the original plant already satisfies $\partial \m{g}/\partial \m{u} \equiv \m{0}$, the pre-filter is not required and the static QP construction applies directly.

From here on we develop the theory on any plant in standard form~\eqref{eq:std_form}, with the supervisory input denoted generically by $\m{u} \in \mathbb{R}^{n_u}$. For models originally in~\eqref{eq:DAE} with $\partial \m{g}/\partial \m{u} \equiv \m{0}$ the theory applies directly; for models with algebraic input dependence it applies to the augmented plant~\eqref{eq:aug_DAE} with $\m{u}$ in~\eqref{eq:std_form} identified with $\m{\nu}$.

\subsection{Time Derivatives Along DAE Trajectories}\label{subsec:hdot}

With the structural form fixed, the next step is to compute how a candidate safety function evolves along DAE trajectories. The key point is that the algebraic variables are not independent states. Their time variation is induced by the differential dynamics and by the disturbance.

For a $C^1$ candidate barrier $h(\m{x}_d, \m{x}_a)$, the time derivative along trajectories of~\eqref{eq:std_form} is
\begin{equation}\label{eq:hdot_chain}
    \dot{h} = \frac{\partial h}{\partial \m{x}_d}\dot{\m{x}}_d + \frac{\partial h}{\partial \m{x}_a}\dot{\m{x}}_a.
\end{equation}
The following lemma makes the algebraic part of this dependence explicit.

\begin{mylem}[Algebraic State Evolution]\label{lem:xa_dot}
Under Assumption~\ref{assump:main}\textit{(i)} applied to~\eqref{eq:std_form}, along any absolutely continuous trajectory of~\eqref{eq:std_form},
\begin{equation}\label{eq:xa_dot}
    \dot{\m{x}}_a = -\m{J}_a^{-1}\!\left[\m{J}_d\,\dot{\m{x}}_d + \frac{\partial \m{g}}{\partial \m{w}}\dot{\m{w}}\right],
\end{equation}
where $\m{J}_a = \partial \m{g}/\partial \m{x}_a$ and $\m{J}_d = \partial \m{g}/\partial \m{x}_d$.
\end{mylem}

\begin{proof}
Differentiating $\m{g}(\m{x}_d(t),\m{x}_a(t),\m{w}(t))=\m{0}$ and using $\partial \m{g}/\partial \m{u}=\m{0}$ gives $\m{J}_d\dot{\m{x}}_d+\m{J}_a\dot{\m{x}}_a+(\partial\m{g}/\partial\m{w})\dot{\m{w}}=\m{0}$. Inverting $\m{J}_a$ gives~\eqref{eq:xa_dot}.
\end{proof}

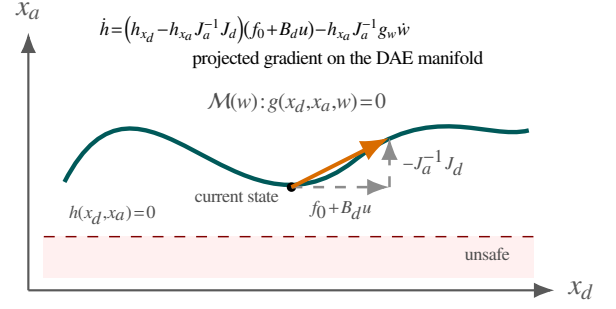
\begin{figure}[t]
\centering
\resizebox{0.9\columnwidth}{!}{%
\begin{tikzpicture}[
  >=Latex,
  font=\footnotesize,
  axis/.style={->, black!65, line width=0.6pt},
  manifold/.style={teal!70!black, line width=1.2pt},
  flow/.style={-Latex, orange!85!black, line width=1.1pt},
  aux/.style={black!45, dashed, line width=0.7pt},
  lab/.style={font=\tiny, text=black!70},
]
  % axes
  \draw[axis] (0,0) -- (5.1,0) node[right] {$x_d$};
  \draw[axis] (0,0) -- (0,2.5) node[above] {$x_a$};

  % unsafe/safe contour bands
  \fill[red!6] (0.15,0.12) rectangle (4.9,0.52);
  \draw[red!50!black, dashed] (0.15,0.52) -- (4.9,0.52)
    node[pos=0.14, above, lab, font=\tiny] {$h(x_d,x_a)=0$};
  \node[red!55!black, lab] at (4.45,0.35) {unsafe};

  % manifold
  \draw[manifold]
    (0.35,1.05) .. controls (1.1,2.5) and (2.0,0.35) ..
    (3.15,1.25) .. controls (3.9,1.85) and (4.45,1.45) .. (4.85,1.55);
  \node[teal!60!black, lab, font=\fontsize{6}{6}\selectfont] at (2.6,1.82)
    {$\mathcal{M}(w):\,g(x_d,x_a,w)=0$};

  % operating point
  \coordinate (p) at (2.55,1);
  \fill[black] (p) circle (1.3pt);
  \node[lab, below left= -0.8mm and 0mm of p, font = \tiny] {current state};

  % differential-only push
  \draw[aux, -Latex] (p) -- ++(0.95,0.0)
    node[midway, below, lab] {$f_0+B_du$};

  % algebraic correction
  \draw[aux, -Latex] ($(p)+(0.95,0)$) -- ++(0.0,0.48)
    node[midway, right, lab] {$-J_a^{-1}J_d$};

  % tangent DAE flow
  \draw[flow] (p) -- ++(0.95,0.48);
    % node[pos=0.65, above, lab] {DAE-consistent direction};

  % formula box
  \node[
    rounded corners=1.2mm,
    % fill=black!4,
    % draw=black!15,
    align=left,
    inner sep=4pt,
    transform shape,
    scale=0.75,
    font=\scriptsize
  ] at (2.55,2.41)
  {$
    \dot h =
    \Bigl(h_{x_d}-h_{x_a}J_a^{-1}J_d\Bigr)
    \bigl(f_0+B_du\bigr)
    -h_{x_a}J_a^{-1}g_w\dot w
  $\\[0.5mm]
  \hspace{12mm}projected gradient on the DAE manifold};

\end{tikzpicture}}
\caption{DAE-consistent barrier differentiation. The algebraic variables are constrained to the manifold \(g(x_d,x_a,w)=0\), so differentiating a safety output requires resolving the induced algebraic motion through \(J_a^{-1}\). This is the origin of the manifold-projected gradient used in the DAE-HOCBF construction.}
\label{fig:dae_projection}
\end{figure}

Substituting~\eqref{eq:xa_dot} and $\dot{\m{x}}_d = \m{f}_0 + \m{B}_d\m{u}$ into~\eqref{eq:hdot_chain} yields the first-order decomposition
\begin{equation}\label{eq:hdot_decomp}
    \dot{h} = a_1(\m{x},\m{w}) + \m{b}_1(\m{x},\m{w})^\top\m{u} + d_{w,1}(\m{x},\m{w},\dot{\m{w}}),
\end{equation}
with
\begin{equation}\label{eq:abd_compact}
    a_1 = \m{p}^\top \m{f}_0,\quad \m{b}_1^\top = \m{p}^\top \m{B}_d,\quad d_{w,1} = -\frac{\partial h}{\partial \m{x}_a}\m{J}_a^{-1}\frac{\partial \m{g}}{\partial \m{w}}\dot{\m{w}},
\end{equation}
and the \emph{manifold-projected gradient}
\begin{equation}\label{eq:proj_grad}
    \m{p}(\m{x},\m{w})^\top := \frac{\partial h}{\partial \m{x}_d} - \frac{\partial h}{\partial \m{x}_a}\m{J}_a^{-1}\m{J}_d.
\end{equation}
Geometrically, $\m{p}(\m{x},\m{w})^\top$ is the gradient of $h$ restricted to the constraint manifold $\mathcal{M}(\m{w}) = \{(\m{x}_d,\m{x}_a):\m{g}(\m{x}_d,\m{x}_a,\m{w})=\m{0}\}$. Figure~\ref{fig:dae_projection} illustrates this DAE-consistent differentiation. A perturbation in the differential state induces an algebraic correction through $\m{J}_a^{-1}$, so the barrier derivative must use the projected gradient on the constraint manifold rather than the ordinary Euclidean gradient.

\subsection{Relative Degree on the DAE}\label{subsec:rel_deg}

Once the first derivative is understood, we can define relative degree consistently with the reduced dynamics on the constraint manifold. This matters in power system models because the same safety output can have different relative degree depending on the surrounding controller stack.

By Assumption~\ref{assump:main}\textit{(i)}, on $\mathcal{M}(\m{w})$ the algebraic states resolve as $\m{x}_a = \m{\varphi}(\m{x}_d,\m{w})$, and the reduced ODE on the manifold reads
\begin{equation}\label{eq:reduced_ODE}
    \dot{\m{x}}_d = \m{F}(\m{x}_d,\m{u},\m{w}) = \m{F}_0(\m{x}_d,\m{w}) + \m{B}(\m{x}_d,\m{w})\m{u},
\end{equation}
with $\m{F}_0(\m{x}_d,\m{w}) = \m{f}_0(\m{x}_d,\m{\varphi}(\m{x}_d,\m{w}),\m{w})$ and $\m{B}(\m{x}_d,\m{w}) = \m{B}_d(\m{x}_d,\m{\varphi}(\m{x}_d,\m{w}),\m{w})$. Let $\m{b}^{(j)}(\m{x}_d,\m{w})$ denote the $j$th column of $\m{B}(\m{x}_d,\m{w})$. The pullback
\begin{equation}\label{eq:H_pullback}
    H(\m{x}_d,\m{w}) := h\bigl(\m{x}_d,\m{\varphi}(\m{x}_d,\m{w})\bigr)
\end{equation}
satisfies $\partial H/\partial \m{x}_d = \m{p}(\m{x},\m{w})^\top$, via $\partial \m{\varphi}/\partial \m{x}_d = -\m{J}_a^{-1}\m{J}_d$.

Repeated differentiation of $H(\m{x}_d,\m{w})$ along a time-varying disturbance introduces the disturbance derivatives $\m{w},\dot{\m{w}},\ldots$ To track this bookkeeping without expanding each derivative term by term, for any integer $q\geq 1$ define the $q$-lifted reduced coordinates
\begin{equation}\label{eq:zeta_q}
    \m{\zeta}_q := \bigl[\m{x}_d^\top,\m{\eta}_0^\top,\m{\eta}_1^\top,\ldots,\m{\eta}_q^\top\bigr]^\top, \qquad \bar H_q(\m{\zeta}_q) := H(\m{x}_d,\m{\eta}_0),
\end{equation}
and the associated frozen-top lifted reduced system
\begin{subequations}\label{eq:lifted_red}
\begin{align}
    \dot{\m{x}}_d &= \m{F}_0(\m{x}_d,\m{\eta}_0) + \m{B}(\m{x}_d,\m{\eta}_0)\m{u}, \label{eq:lifted_red_a}\\
    \dot{\m{\eta}}_\ell &= \m{\eta}_{\ell+1}, \qquad \ell = 0,\ldots,q-1, \label{eq:lifted_red_b}\\
    \dot{\m{\eta}}_q &= \m{0}. \label{eq:lifted_red_c}
\end{align}
\end{subequations}
The terminal equation is a bookkeeping device only. Derivatives of $\bar H_q$ up to order $q$ depend on $\m{\eta}_q$ but not on $\dot{\m{\eta}}_q$, so freezing the top lifted disturbance derivative does not change the expressions needed below. For each input component $u_j$, define the lifted input vector field
\begin{equation}\label{eq:lifted_b}
    \bar{\m{b}}_q^{(j)}(\m{\zeta}_q) := \bigl[(\m{b}^{(j)}(\m{x}_d,\m{\eta}_0))^\top,\m{0}^\top,\ldots,\m{0}^\top\bigr]^\top.
\end{equation}
The following definition formalizes the first derivative order at which the supervisory input enters the barrier dynamics.

\begin{mydef}[Relative Degree]\label{def:rel_deg}
Let $\widehat{\mathcal{D}}_q\subset\mathbb{R}^{n_d + (q+1)n_w}$ be open in the lifted coordinates $\m{\zeta}_q$, and let $\mathcal{D}$ denote the associated manifold domain obtained by identifying $\m{\eta}_0=\m{w}$ and $\m{x}_a=\m{\varphi}(\m{x}_d,\m{w})$. The \emph{relative degree} of $h$ w.r.t.\ $u_j$ is the smallest integer $r_j \geq 1$ such that, on some open lifted domain $\widehat{\mathcal{D}}_{r_j}$,
\begin{equation}\label{eq:rel_deg_Lie}
\begin{split}
    &L_{\bar{\m{b}}_{r_j}^{(j)}} L_{\bar{\m{F}}_{0,r_j}}^{\,r_j-1} \bar H_{r_j}(\m{\zeta}_{r_j}) \neq 0 \; \text{ on }\widehat{\mathcal{D}}_{r_j}, \\ L_{\bar{\m{b}}_{r_j}^{(j)}} &L_{\bar{\m{F}}_{0,r_j}}^{\,\ell-1} \bar H_{r_j} \equiv 0 \; \text{ on }\widehat{\mathcal{D}}_{r_j}, \;\; \ell = 1,\ldots,r_j-1,
    \end{split}
\end{equation}
where $\bar{\m{F}}_{0,q}$ is the drift of~\eqref{eq:lifted_red} and $L_{\m{X}}\phi = (\partial \phi/\partial \m{\zeta}_q)\m{X}$. The relative degree of $h$ w.r.t.\ the vector $\m{u}$ is $r := \min_j r_j$, with $\widehat{\mathcal{D}}_r$ the common lifted domain on which~\eqref{eq:rel_deg_Lie} holds and $\mathcal{D}$ its associated manifold domain.
\end{mydef}
When $r_{j_1}=r_{j_2}=r$ for distinct components $j_1,j_2$, the resulting $\m{B}_r$ has nonzero entries in those components and the multi-input coupling of Remark~\ref{rem:multi_input} applies; otherwise $\m{B}_r$ is supported on the single component achieving the minimum relative degree. The uniformity clause, namely that the vanishing and non-vanishing conditions hold pointwise on an open set in the joint variables $(\m{x}_d,\m{w},\dot{\m{w}},\ldots)$, is required for the HOCBF construction below, for the same reason as in~\cite[Def.~7]{xiao2021high}.

The first lemma is the standard iterated Lie derivative identity for control-affine ODEs~\cite[Ch.~4]{isidori1985nonlinear}. We state it explicitly because it is the template for the DAE calculation below.

\begin{mylem}[Iterated Lie Derivatives, Constant Disturbance]\label{lem:ode_chain}
Consider $\dot{\m{x}}_d = \m{F}_0(\m{x}_d) + \m{B}(\m{x}_d)\m{u}$ on $\mathcal{D}\subset\mathbb{R}^{n_d}$. Let $H$ be $C^r$ and have relative degree $r\geq 1$ w.r.t.\ $\m{u}$. Along any absolutely continuous trajectory with measurable, bounded input, the following identities hold for a.e.\ $t$:
\begin{enumerate}[label=(\roman*)]
    \item $H^{(k)}(t) = L_{\m{F}_0}^k H(\m{x}_d(t))$ for a.e.\ $t$ and $k = 0,\ldots,r-1$, so these derivatives do not depend on $\m{u}$ or on derivatives of $\m{u}$.
    \item For a.e.\ $t$, the first input-dependent derivative is
    \[
        H^{(r)}(t) = L_{\m{F}_0}^r H + \sum_{j=1}^{n_u} u_j(t)L_{\m{b}^{(j)}} L_{\m{F}_0}^{r-1} H,
    \]
    evaluated at $\m{x}_d(t)$, and at least one coefficient of $u_j$ is nonzero on $\mathcal{D}$.
\end{enumerate}
\end{mylem}

\begin{proof}
We prove \textit{(i)} by induction on $k$. The base case $k=0$ is tautological, since $H^{(0)} = H$ depends on $\m{x}_d$ only.

Inductive step: suppose $H^{(k)}(\m{x}_d)$ is a smooth function of $\m{x}_d$ alone for some $0 \leq k \leq r-2$. Then
\begin{equation}\label{eq:one_step}
\begin{aligned}
H^{(k+1)}(t)
&= \frac{d}{dt}H^{(k)}(\m{x}_d(t)) \\
&= \frac{\partial H^{(k)}}{\partial \m{x}_d}
   \bigl[\m{F}_0+\m{B}\m{u}(t)\bigr] \\
&= L_{\m{F}_0}H^{(k)}
   +\sum_{j=1}^{n_u}u_j(t)L_{\m{b}^{(j)}}H^{(k)} .
\end{aligned}
\end{equation}
Because $H^{(k)}$ depends only on $\m{x}_d$, no $\dot{\m{u}}$ appears in~\eqref{eq:one_step}. Applying the induction hypothesis once more, $H^{(k)} = L_{\m{F}_0}^k H$, and so $L_{\m{b}^{(j)}} H^{(k)} = L_{\m{b}^{(j)}} L_{\m{F}_0}^k H$. For $k+1 \leq r-1$, that is $k \leq r-2$, the vanishing half of the ODE relative degree condition gives $L_{\m{b}^{(j)}} L_{\m{F}_0}^k H \equiv 0$ on $\mathcal{D}$ for every $j$, so~\eqref{eq:one_step} reduces to $H^{(k+1)} = L_{\m{F}_0}^{k+1} H$, a function of $\m{x}_d$ alone, which closes the induction.

For \textit{(ii)}, setting $k+1 = r$ in~\eqref{eq:one_step} with $H^{(r-1)} = L_{\m{F}_0}^{r-1} H$ from \textit{(i)} gives $H^{(r)} = L_{\m{F}_0}^r H + \sum_j u_j L_{\m{b}^{(j)}} L_{\m{F}_0}^{r-1} H$. The non-vanishing half of the ODE relative degree condition provides at least one nonzero coefficient $L_{\m{b}^{(j)}} L_{\m{F}_0}^{r-1} H$ on $\mathcal{D}$.
\end{proof}

\begin{figure}[t]
\centering
\resizebox{\columnwidth}{!}{%
\begin{tikzpicture}[
  >=Latex,
  font=\footnotesize,
  node distance=4mm and 5mm,
  box/.style={rounded corners=1.5mm, draw=none, fill=black!4,
              align=center, inner sep=4pt, minimum height=7mm},
  eta/.style={box, fill=gray!8, minimum width=12mm},
  calc/.style={box, fill=teal!12, minimum width=18mm},
  qprow/.style={box, fill=orange!12, minimum width=32mm},
  arr/.style={-Latex, line width=0.9pt, black!70},
  note/.style={font=\scriptsize, text=black!60, align=center}
]
  % exosystem chain
  \node[eta] (e0) {$\eta_0=w$};
  \node[eta, right=of e0] (e1) {$\eta_1=\dot w$};
  \node[eta, right=of e1] (e2) {$\cdots$};
  \node[eta, right=of e2] (er1) {$\eta_{r-1}$};
  \node[eta, right=of er1] (er) {$\eta_r\in\mathcal V$};

  \draw[arr] (e0) -- (e1);
  \draw[arr] (e1) -- (e2);
  \draw[arr] (e2) -- (er1);
  \draw[arr] (er1) -- (er);

  \node[note, below=1mm of e1, xshift=19mm, yshift=-3mm] {disturbance\\derivative states};
  \node[note, below=1mm of er, xshift=-3mm, yshift=-4mm] {bounded top\\derivative};

  % manifold and derivative engine
  \node[calc, below=11mm of e0] (manifold)
  {DAE manifold\\[0.2mm] \(g(x_d,x_a,w)=0\)};
  \node[calc, right=7mm of manifold] (lift)
  {lifted Lie\\derivatives};
  \node[qprow, right=7mm of lift] (row)
  {\(\displaystyle A_r+B_r^\top u+\Gamma_r^\top\eta_r+\pi_{r-1}\ge 0\)};

  \draw[arr] (e0.south) -- (manifold.north);
  \draw[arr] (e1.south) -| ++ (0mm,-5mm) -| (lift.north);
  \draw[arr] (er.south) -| ++ (0mm,-5mm) -| (row.north);
  \draw[arr] (manifold) -- (lift);
  \draw[arr] (lift) -- (row);

  % robustification
  \node[qprow, below=7mm of row] (robust)
  {\(\displaystyle A_r+B_r^\top u+\underline D_r+\pi_{r-1}\ge 0\)};
  \draw[arr] (row) -- node[right=1mm, note] {worst case\\top derivative} (robust);

  \node[note,  below left =5.5mm and -25mm of lift]
  {For \(k<r\), \(h^{(k)}=\Phi_k(x_d,w^{[k]})\) is input free.\\
   At \(k=r\), the derivative is affine in \(u\) and linear in \(w^{(r)}\).};

\end{tikzpicture}}
\caption{Lifted disturbance bookkeeping and robust HOCBF row construction. The disturbance derivatives are represented by an integrator chain exosystem. Below relative degree, the barrier derivatives are independent of the supervisory input; at relative degree \(r\), the expression becomes affine in \(u\) and linear in the top disturbance derivative, yielding a robust affine QP constraint.}
\label{fig:lifted_hocbf_row}
\end{figure}
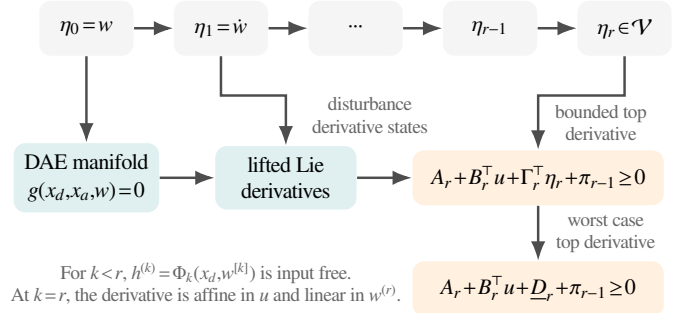

Figure~\ref{fig:lifted_hocbf_row} summarizes the construction. The exosystem tracks the derivatives of \(w\), the lower order barrier derivatives stay input free, and the derivative at relative degree \(r\) becomes an affine robust QP constraint in the supervisory input.

The next lemma lifts this statement to the DAE with time-varying disturbances. It supplies the derivative form used later to build the HOCBF row.

\begin{mylem}[DAE-Lift with Time-Varying Disturbance]\label{lem:dae_lift}
Consider the standard form DAE~\eqref{eq:std_form} under Assumption~\ref{assump:main}. Let $h$ have uniform relative degree $r\geq 1$ w.r.t.\ $\m{u}$ on $\widehat{\mathcal{D}}_r$, with associated manifold domain $\mathcal{D}$. Along trajectories that stay in $\mathcal{D}$, the derivatives of $h$ have the following form. For $\m{w}^{[\ell]} := (\m{w},\dot{\m{w}},\ldots,\m{w}^{(\ell)})$, there exist smooth maps
\[
    \Phi_k: \pi_{\m{x}_d}(\mathcal{D})\times\mathbb{R}^{(k+1)\,n_w}\to\mathbb{R}, \quad k=0,1,\ldots,r-1,
\]
and smooth maps $A_r$, $\m{B}_r$, and $D_r$ of compatible dimensions such that, for any $\m{w}(\cdot)\in\mathcal{W}_r$, the identities below hold for a.e.\ $t$:
\begin{subequations}\label{eq:dae_lift}
\begin{align}
    h^{(k)}(t) &= \Phi_k\bigl(\m{x}_d(t),\m{w}^{[k]}(t)\bigr), \quad k=0,\ldots,r-1, \label{eq:dae_lift_below}\\
    h^{(r)}(t) &= A_r\bigl(\m{x}_d(t),\m{w}^{[r-1]}(t)\bigr) + \m{B}_r\bigl(\m{x}_d(t),\m{w}^{[r-1]}(t)\bigr)^{\!\top}\m{u}(t) \notag \\ &+ D_r\bigl(\m{x}_d(t),\m{w}^{[r]}(t)\bigr). \label{eq:dae_lift_at_r}
\end{align}
\end{subequations}
Thus the derivatives below relative degree are input free, and the derivative at relative degree is affine in the supervisory input. More precisely:
\begin{enumerate}[label=(\roman*)]
    \item $h^{(k)}$ is independent of $\m{u}$, $\m{x}_a$, and all derivatives of $\m{u}$ for $k=0,\ldots,r-1$.
    \item $\m{B}_r$ is nonzero on $\widehat{\mathcal{D}}_r$. On the constant disturbance slice $\dot{\m{w}}=\cdots=\m{w}^{(r-1)}=\m{0}$,
    \[
        \m{B}_r(\m{x}_d,\m{w},\m{0},\ldots,\m{0}) = \bigl[L_{\m{b}^{(1)}} L_{\m{F}_0}^{r-1} H,\ldots,L_{\m{b}^{(n_u)}} L_{\m{F}_0}^{r-1} H\bigr]^\top(\m{x}_d,\m{w});
    \]
    $A_r$ and $D_r$ are $\m{u}$-independent, and $D_r$ is linear in the top disturbance derivative $\m{w}^{(r)}$.
    \item No derivative of $\m{u}$ appears in~\eqref{eq:dae_lift}.
\end{enumerate}
\end{mylem}

\begin{proof}
On $\mathcal{M}(\m{w})$, $\m{x}_a = \m{\varphi}(\m{x}_d,\m{w})$ and $h(\m{x}) = H(\m{x}_d,\m{w})$ via~\eqref{eq:H_pullback}. Fix the $r$-lifted reduced system~\eqref{eq:lifted_red} with lifted state $\m{\zeta}_r = [\m{x}_d^\top,\m{\eta}_0^\top,\ldots,\m{\eta}_r^\top]^\top$ and lifted output $\bar H_r(\m{\zeta}_r)=H(\m{x}_d,\m{\eta}_0)$. By Definition~\ref{def:rel_deg}, $\bar H_r$ has relative degree $r$ w.r.t.\ $\m{u}$ on the lifted domain $\widehat{\mathcal{D}}_r$ corresponding to $\mathcal{D}$.

Applying Lemma~\ref{lem:ode_chain} to the lifted control-affine ODE gives
\begin{equation}\label{eq:lifted_chain_below}
    \frac{d^k}{dt^k}\bar H_r(\m{\zeta}_r(t)) = L_{\bar{\m{F}}_{0,r}}^k \bar H_r(\m{\zeta}_r(t)), \qquad k=0,\ldots,r-1,
\end{equation}
and
\begin{equation}\label{eq:lifted_chain_at_r}
    \frac{d^r}{dt^r}\bar H_r(\m{\zeta}_r(t)) = L_{\bar{\m{F}}_{0,r}}^r \bar H_r(\m{\zeta}_r(t)) + \sum_{j=1}^{n_u} u_j(t)\,L_{\bar{\m{b}}_r^{(j)}}L_{\bar{\m{F}}_{0,r}}^{\,r-1}\bar H_r(\m{\zeta}_r(t)).
\end{equation}
Now evaluate these expressions along the actual disturbance history by setting $\m{\eta}_\ell(t)=\m{w}^{(\ell)}(t)$ for $\ell=0,\ldots,r$. Since $\bar H_r$ depends on the lifted coordinates only through $(\m{x}_d,\m{\eta}_0)$ and the lifted drift propagates the disturbance chain one level per differentiation, $L_{\bar{\m{F}}_{0,r}}^k \bar H_r$ depends only on $(\m{x}_d,\m{\eta}_0,\ldots,\m{\eta}_k)$ for $k\leq r$. Hence there exist smooth maps $\Phi_k$ such that $h^{(k)}(t) = \Phi_k(\m{x}_d(t),\m{w}^{[k]}(t))$ for $k=0,\ldots,r-1$, which establishes~\eqref{eq:dae_lift_below} and proves part \textit{(i)}.

For the derivative of order $r$, define
\begin{align}
    \m{B}_r(\m{x}_d,\m{w}^{[r-1]}) :=& \bigl[L_{\bar{\m{b}}_r^{(1)}}L_{\bar{\m{F}}_{0,r}}^{\,r-1}\bar H_r,\ldots, L_{\bar{\m{b}}_r^{(n_u)}}L_{\bar{\m{F}}_{0,r}}^{\,r-1}\bar H_r\bigr]^\top\!\Big|_{\substack{\m{\eta}_0=\m{w}\\ \m{\eta}_\ell=\m{w}^{(\ell)},\,\ell=1,\ldots,r-1}}, \label{eq:Br_def_new}\\
    A_r(\m{x}_d,\m{w}^{[r-1]}) :=& \Bigl(L_{\bar{\m{F}}_{0,r}}^r \bar H_r\Bigr)\Big|_{\substack{\m{\eta}_0=\m{w},\,\ldots,\,\m{\eta}_{r-1}=\m{w}^{(r-1)}\\ \m{\eta}_r=\m{0}}}, \label{eq:Ar_def}\\
    D_r(\m{x}_d,\m{w}^{[r]}) :=& \Bigl(L_{\bar{\m{F}}_{0,r}}^r \bar H_r\Bigr)\Big|_{\substack{\m{\eta}_0=\m{w},\,\ldots,\,\m{\eta}_r=\m{w}^{(r)}}} - \Bigl(L_{\bar{\m{F}}_{0,r}}^r \bar H_r\Bigr)\Big|_{\substack{\m{\eta}_0=\m{w},\,\ldots,\,\m{\eta}_{r-1}=\m{w}^{(r-1)}\\ \m{\eta}_r=\m{0}}}. \label{eq:Dr_def}
\end{align}
Because $\m{\eta}_r$ enters the lifted drift only through the single integrator equation $\dot{\m{\eta}}_{r-1}=\m{\eta}_r$, the quantity $L_{\bar{\m{F}}_{0,r}}^r \bar H_r$ is affine in $\m{\eta}_r$, so $D_r$ is linear in $\m{w}^{(r)}$. Substituting~\eqref{eq:Ar_def} and~\eqref{eq:Dr_def} into~\eqref{eq:lifted_chain_at_r} yields~\eqref{eq:dae_lift_at_r}. The coefficient $\m{B}_r$ is nonzero on $\widehat{\mathcal{D}}_r$ by the non-vanishing half of Definition~\ref{def:rel_deg}, which establishes part \textit{(ii)}. On the constant disturbance slice $\m{\eta}_1=\cdots=\m{\eta}_{r-1}=\m{0}$ the lifted system reduces to the frozen disturbance reduced ODE, so
\[
    \m{B}_r(\m{x}_d,\m{w},\m{0},\ldots,\m{0}) = \bigl[L_{\m{b}^{(1)}} L_{\m{F}_0}^{r-1} H,\ldots,L_{\m{b}^{(n_u)}} L_{\m{F}_0}^{r-1} H\bigr]^\top(\m{x}_d,\m{w}),
\]
exactly as claimed. Part \textit{(iii)} follows from~\eqref{eq:lifted_chain_below} and~\eqref{eq:lifted_chain_at_r}, since the control enters only through $\m{u}$ itself and never through its derivatives.
\end{proof}

Together, the two lemmas give the chain rule statement used by the barrier construction.

\begin{mycor}[Chain Rule Expansion]\label{cor:chain_rule}
Under the hypotheses of Lemma~\ref{lem:dae_lift}, the DAE derivatives of $h$ are input free up to order $r-1$. At order $r$ they have the affine form~\eqref{eq:dae_lift_at_r}, with $\m{B}_r\not\equiv\m{0}$ on $\widehat{\mathcal{D}}_r$, $D_r$ linear in $\m{w}^{(r)}$, and no derivatives of $\m{u}$.
\end{mycor}

\begin{proof}
Direct from Lemmas~\ref{lem:ode_chain} and~\ref{lem:dae_lift}.
\end{proof}

Since $D_r$ is linear in the top disturbance derivative, we write
\[ D_r(\m{x}_d,\m{w}^{[r]})=\m{\Gamma}_r(\m{x}_d,\m{w}^{[r-1]})^\top \m{w}^{(r)} \]
for the corresponding smooth coefficient map
$\m{\Gamma}_r:\pi_{\m{x}_d}(\mathcal{D})\times\mathbb{R}^{r n_w}\to\mathbb{R}^{n_w}$.
For $k > r$, differentiating $\m{B}_r^\top \m{u}$ would produce $\dot{\m{u}}$, but the HOCBF construction below never expands beyond order $r$, so this causes no difficulty. The $r=1$ case is included without modification in the lifted-system proof of Lemma~\ref{lem:dae_lift}.

The construction is independent of the internal order of any particular controller block. Only the state partition, the supervisory input, and the relative degree of each safety output change. Additional controller states between the supervisory input and the protected variable generally raise the relative degree, and when the pre-filter of Section~\ref{subsec:prefilter} is used, the relative degree w.r.t.\ the new command is one higher than w.r.t.\ the pre-filter output state in the augmented model.

\subsection{DAE-HOCBF, Feasibility, and Forward Invariance}\label{subsec:gen_HOCBF}

We now turn the derivative formulas into enforceable higher order barrier conditions. We first define the barrier recursion, then give three guarantees: exact row feasibility, hard forward invariance, and a computable margin when the last barrier row is violated by a bounded amount.

We adopt the HOCBF recursion of~\cite{xiao2021high}, adapted to the DAE setting. Let $h$ have relative degree $r$ on the lifted domain $\widehat{\mathcal{D}}_r$ of Definition~\ref{def:rel_deg}, with associated manifold domain $\mathcal{D}$, and let $\alpha_1,\ldots,\alpha_r$ be extended class-$\mathcal{K}$ functions with $\alpha_k$ being $(r-k)$-times continuously differentiable. Define
\begin{subequations}\label{eq:psi_seq}
\begin{align}
    &\psi_0 := h, \\
    &\psi_k := \dot\psi_{k-1} + \alpha_k(\psi_{k-1}), \quad k = 1,\ldots,r,
\end{align}
\end{subequations}
and the nested safe sets $\mathcal{C}_k := \{\m{x}:\psi_k \geq 0\}$ for $k=0,\ldots,r-1$. By Corollary~\ref{cor:chain_rule}, $\psi_0,\ldots,\psi_{r-1}$ are independent of $\m{u}$, while $\psi_r$ is affine in $\m{u}$ and depends smoothly on $(\m{x},\m{w}^{[r]})$. Expanding $\psi_r$,
\begin{equation}\label{eq:psi_r_affine}
\begin{split}
    \psi_r(\m{x},\m{u},\m{w}^{[r]}) &= A_r(\m{x},\m{w}^{[r-1]}) + \m{B}_r(\m{x},\m{w}^{[r-1]})^\top \m{u} \\ &+ D_r(\m{x},\m{w}^{[r]}) + \pi_{r-1}(\m{x},\m{w}^{[r-1]}),
    \end{split}
\end{equation}
where $\pi_{r-1}$ collects all remaining input-free terms in the HOCBF recursion, including the terminal term $\alpha_r(\psi_{r-1})$ and the lower order terms generated by $\alpha_1(\psi_0),\ldots,\alpha_{r-1}(\psi_{r-2})$. Thus $\pi_{r-1}$ depends on $(\m{x},\m{w}^{[r-1]})$ through the $\Phi_k$-recursion of Lemma~\ref{lem:dae_lift}, but not on $\m{u}$ or on derivatives of $\m{u}$.

We first isolate the structural data of a barrier and its recursion, then add the feasibility requirement that makes it enforceable.

\begin{mydef}[Relative Degree $r$ Barrier Candidate]\label{def:barrier_candidate}
A \emph{relative degree $r$ barrier candidate} for~\eqref{eq:std_form} is a function $h\in C^r$ with uniform relative degree $r$ on $\widehat{\mathcal{D}}_r$, together with extended class-$\mathcal{K}$ functions $\alpha_1,\ldots,\alpha_r$ such that $\alpha_k$ is $(r-k)$-times continuously differentiable. The recursion $\psi_0,\ldots,\psi_r$ is defined by~\eqref{eq:psi_seq}. We set $\mathcal{C}_k := \{\m{x}:\psi_k\geq 0\}$ for $k=0,\ldots,r-1$ and $\mathcal{C}_{\cap} := \bigcap_{k=0}^{r-1}\mathcal{C}_k \cap \mathcal{D}$.
\end{mydef}

Definition~\ref{def:barrier_candidate} gives the smooth function, relative degree, and gains. It does not yet say that an admissible input can enforce the last row $\psi_r\geq0$.

\begin{mydef}[DAE-HOCBF]\label{def:DAE_HOCBF}
A relative degree $r$ barrier candidate $h$ is a \emph{DAE-HOCBF} on $\mathcal{C}_{\cap}$ if the robust admissible input set is nonempty at every point of interest. That set is
\begin{equation}\label{eq:adm_set}
    \mathcal{U}_h(\m{x},\m{w}^{[r-1]}) := \bigl\{\m{u}\in\mathcal{U} \;\big|\; A_r + \m{B}_r^\top\m{u} + \underline{D}_r + \pi_{r-1} \geq 0\bigr\}
\end{equation}
for every $\m{x}\in\mathcal{C}_{\cap}$ on $\mathcal{M}(\m{w})$ and every admissible disturbance history. Here $A_r, \m{B}_r, \pi_{r-1}$ are as in~\eqref{eq:psi_r_affine}, $\mathcal{U} := [\m{u}_{\min},\m{u}_{\max}]$, and
\begin{equation}\label{eq:underline_Dr}
    \underline{D}_r(\m{x},\m{w}^{[r-1]}) := \inf_{\|\m{w}^{(r)}\|\leq \bar{w}_r} D_r(\m{x},\m{w}^{[r-1]},\m{w}^{(r)})
\end{equation}
is the worst case contribution of the top disturbance derivative. Since $D_r=\m{\Gamma}_r^\top \m{w}^{(r)}$, Lemma~\ref{lem:dae_lift} gives $\underline{D}_r = -\bar{w}_r\,\|\m{\Gamma}_r\|_*$, where $\|\cdot\|_*$ is the dual norm.
\end{mydef}

If the admissible input set is empty at some operating point, hard safety cannot be enforced without additional actuation authority, and the soft QP of Section~\ref{subsec:QP} uses slack. For compactness we write the pointwise condition as
\begin{equation}\label{eq:HOCBF_cond}
\begin{split}
    A_r(\m{x},\m{w}^{[r-1]}) &+ \m{B}_r(\m{x},\m{w}^{[r-1]})^\top \m{u} \\&+ \underline{D}_r(\m{x},\m{w}^{[r-1]}) + \pi_{r-1}(\m{x},\m{w}^{[r-1]}) \geq 0.
\end{split}
\end{equation}

For box constrained inputs, nonemptiness reduces to one scalar residual. This residual is used online to detect whether a hard row is enforceable and offline to separate the hard safety regime from the authority limited regime.

\begin{theorem}[Exact Robust Feasibility of One DAE-HOCBF Row]\label{thm:robust_row_feasibility}
Fix a relative degree $r$ barrier candidate and a point $(\m{x},\m{w}^{[r-1]})$ with $\m{x}\in\mathcal{C}_{\cap}$ on $\mathcal{M}(\m{w})$. Define
\begin{equation}\label{eq:rho_h_def}
    \begin{aligned}
    \sigma_{\mathcal{U}}(\m{B}_r)&:=\sup_{\m{u}\in\mathcal{U}}\m{B}_r^\top\m{u}
    =\m{B}_r^\top\bar{\m{u}}+\tfrac12|\m{B}_r|^\top\Delta\m{u},\\
    \rho_h&:=A_r+\pi_{r-1}+\underline{D}_r+\sigma_{\mathcal{U}}(\m{B}_r),
    \end{aligned}
\end{equation}
where $\bar{\m{u}}:=\tfrac12(\m{u}_{\max}+\m{u}_{\min})$, $\Delta\m{u}:=\m{u}_{\max}-\m{u}_{\min}$, and $|\m{B}_r|$ is componentwise. All terms are evaluated at $(\m{x},\m{w}^{[r-1]})$. The robust row~\eqref{eq:adm_set} is feasible over the actuator box if and only if $\rho_h\geq0$.

The maximizing input is the box vertex $\m{u}^{\dagger}$ with $u_i^{\dagger}=u_{\max,i}$ when $(\m{B}_r)_i\geq0$ and $u_i^{\dagger}=u_{\min,i}$ otherwise. If $\rho_h<0$, then there is an admissible top disturbance derivative $\m{v}^{\star}$ that makes the row negative for every $\m{u}\in\mathcal{U}$. For the relaxed row with slack $\xi\geq0$, the least feasible slack is $\xi_{\min}=(-\rho_h)_+$.
\end{theorem}

\begin{proof}
Because $\mathcal{U}$ is a box, $\sup_{u_i\in[u_{\min,i},u_{\max,i}]}(\m{B}_r)_i u_i$ equals $(\m{B}_r)_i u_{\max,i}$ when $(\m{B}_r)_i\geq0$ and $(\m{B}_r)_i u_{\min,i}$ otherwise. Writing $u_{\max,i}=\bar u_i+\tfrac12\Delta u_i$ and $u_{\min,i}=\bar u_i-\tfrac12\Delta u_i$, both cases combine to $(\m{B}_r)_i\bar u_i+\tfrac12|(\m{B}_r)_i|\Delta u_i$. Summing over $i$ gives $\sigma_{\mathcal{U}}(\m{B}_r)$ and the vertex $\m{u}^{\dagger}$ in~\eqref{eq:rho_h_def}.

The largest value of $A_r+\pi_{r-1}+\underline{D}_r+\m{B}_r^\top\m{u}$ over $\m{u}\in\mathcal{U}$ is exactly $\rho_h$. Hence $\mathcal{U}_h$ is nonempty if and only if $\rho_h\geq0$, with $\m{u}^{\dagger}$ as a witness. The map $\m{v}\mapsto\m{\Gamma}_r^\top\m{v}$ is continuous on the compact ball $\{\|\m{v}\|\leq\bar{w}_r\}$, so the infimum defining $\underline{D}_r$ is attained at some $\m{v}^{\star}$. If $\rho_h<0$, then for every $\m{u}\in\mathcal{U}$,
\[
    A_r+\pi_{r-1}+\m{\Gamma}_r^\top\m{v}^{\star}+\m{B}_r^\top\m{u}
    \leq A_r+\pi_{r-1}+\underline{D}_r+\sigma_{\mathcal{U}}(\m{B}_r)
    =\rho_h<0.
\]
Finally, the relaxed row is feasible if and only if $\rho_h+\xi\geq0$, whose least nonnegative solution is $(-\rho_h)_+$.
\end{proof}

The residual $\rho_h$ splits the operating domain in two. Where $\rho_h\geq0$, the hard barrier row is enforceable. Where $\rho_h<0$, the deficit $(-\rho_h)_+$ is unavoidable for that row. The next theorem covers the hard safety regime.

\begin{theorem}[Forward Invariance via DAE-HOCBF]\label{thm:forward_inv}
Let $h$ be a DAE-HOCBF of relative degree $r$ on $\mathcal{C}_{\cap}$. Consider~\eqref{eq:std_form} under Assumption~\ref{assump:main} with $\m{w}(\cdot)\in\mathcal{W}_r$. Suppose the input is measurable, takes values in $\mathcal{U}$, and satisfies
\begin{equation}\label{eq:thm_u_cond}
    \m{u}(t)\in\mathcal{U}_h(\m{x}(t),\m{w}^{[r-1]}(t)) \quad \text{for a.e. } t\geq 0 \text{ such that } \m{x}(t)\in\mathcal{D},
\end{equation}
which is equivalent to enforcing the robust HOCBF row~\eqref{eq:HOCBF_cond}. If the initial condition is algebraically consistent, $\psi_k(0)\geq0$ for $k=0,\ldots,r-1$, and the trajectory remains in $\mathcal{D}$ on $[0,T]$, then
\[
    \psi_k(t)\geq0,\quad k=0,\ldots,r-1,\qquad h(\m{x}(t))\geq0,
\]
for all $t\in[0,T]$. The algebraic constraint also remains satisfied on $[0,T]$.
\end{theorem}

\begin{proof}
By Assumption~\ref{assump:main}\textit{(i)}, the implicit function theorem yields locally on $\mathcal{D}$ a $C^{s-1}$ map $\m{x}_a=\m{\varphi}(\m{x}_d,\m{w})$ describing the consistency manifold. Since the initial state is algebraically consistent and the trajectory stays in $\mathcal{D}$ on $[0,T]$, the solution evolves on that manifold, so $\m{g}(\m{x}(t),\m{w}(t))=\m{0}$ throughout.

For each $k=0,\ldots,r-1$, Corollary~\ref{cor:chain_rule} gives $\psi_k(t)=\Psi_k(\m{x}_d(t),\m{w}^{[k]}(t))$ for some $C^1$ map $\Psi_k$ on a neighborhood of the compact image of $[0,T]$. Because $\m{u}\in L^\infty([0,T];\mathcal{U})$ and $\dot{\m{x}}_d = \m{f}_0 + \m{B}_d\m{u}$ with $\m{f}_0,\m{B}_d$ smooth, $\m{x}_d(\cdot)$ is absolutely continuous on $[0,T]$, and by Assumption~\ref{assump:main}\textit{(iii)} each $\m{w}^{(\ell)}(\cdot)$ is absolutely continuous for $\ell=0,\ldots,r-1$. The stacked signal $t\mapsto (\m{x}_d(t),\m{w}^{[k]}(t))$ is therefore absolutely continuous, and since $\Psi_k$ is locally Lipschitz on compacts, each $\psi_k$ is absolutely continuous on $[0,T]$, so the comparison lemma~\cite[Lem.~4.4]{khalil2002nonlinear} applies.

We prove $\psi_k\geq0$ by downward induction on $k$. For the base case $k=r-1$, condition~\eqref{eq:thm_u_cond} gives $\m{u}(t)\in\mathcal{U}_h$ for a.e.\ $t\in[0,T]$, so~\eqref{eq:HOCBF_cond} holds a.e. By the definition of $\underline{D}_r$ and the bound $\|\m{w}^{(r)}(t)\|\leq \bar{w}_r$, this yields $\psi_r(t)\geq 0$ a.e., that is $\dot{\psi}_{r-1}(t)\geq -\alpha_r(\psi_{r-1}(t))$. With $\psi_{r-1}(0)\geq 0$ and $\alpha_r$ extended class-$\mathcal{K}$, the comparison lemma gives $\psi_{r-1}(t)\geq 0$ on $[0,T]$. For the inductive step, assume $\psi_k(t)\geq 0$ on $[0,T]$ for some $1\leq k\leq r-1$. From~\eqref{eq:psi_seq}, $\dot{\psi}_{k-1}(t)=\psi_k(t)-\alpha_k(\psi_{k-1}(t))\geq -\alpha_k(\psi_{k-1}(t))$ a.e., and with $\psi_{k-1}(0)\geq 0$ the comparison lemma gives $\psi_{k-1}(t)\geq 0$ on $[0,T]$. Descending to $k=0$ gives $\psi_k\geq0$ for all $k$, and $h=\psi_0$ completes the proof.
\end{proof}

Theorem~\ref{thm:forward_inv} requires the last recursion term to be nonnegative. In practice, slack, model error, or sampled-data implementation may leave a bounded violation. The next result converts that violation into a computable safe set inflation.

\begin{theorem}[Input-to-State Safety Margin]\label{thm:dae_issf_margin}
Let $h$ be a relative degree $r$ barrier candidate with recursion~\eqref{eq:psi_seq}. Consider an algebraically consistent trajectory that stays in $\mathcal{D}$ on $[0,T]$, with each $\psi_k$ absolutely continuous. Suppose
\[
    \psi_r(t)\geq-\delta(t)\quad \text{for a.e. }t\in[0,T],
\]
with $\delta\in L^\infty([0,T])$ and $\delta(t)\geq0$. Let $\beta_r\geq0$ satisfy $\delta(t)\leq\beta_r$ for a.e.\ $t\in[0,T]$, and choose constants $\beta_{k-1}\geq0$ satisfying $\alpha_k(-\beta_{k-1})\leq-\beta_k$ for $k=r,\ldots,1$. If $\psi_k(0)\geq-\beta_k$ for $k=0,\ldots,r-1$, then $\psi_k(t)\geq-\beta_k$ on $[0,T]$. In particular, $h(\m{x}(t))\geq-\beta_0$.

When each $\alpha_k$ is onto $\mathbb{R}$, the least such constants are $\beta_{k-1}=-\alpha_k^{-1}(-\beta_k)$. For linear gains $\alpha_k(s)=\gamma_k s$,
\begin{equation}\label{eq:linear_margin}
    \beta_0=\frac{\beta_r}{\prod_{k=1}^{r}\gamma_k},
    \qquad
    h(\m{x}(t))\geq -\beta_0,\quad t\in[0,T].
\end{equation}
\end{theorem}

\begin{proof}
We argue by backward induction through the recursion. The hypothesis gives $\dot{\psi}_{r-1}(t)\geq-\beta_r-\alpha_r(\psi_{r-1}(t))$ a.e. Suppose $\psi_{r-1}(0)\geq-\beta_{r-1}$ and $\alpha_r(-\beta_{r-1})\leq-\beta_r$, and suppose for contradiction that $\psi_{r-1}(t_1)<-\beta_{r-1}$ for some $t_1$. Let $t_0=\sup\{t\leq t_1:\psi_{r-1}(t)\geq-\beta_{r-1}\}$, so by continuity $\psi_{r-1}(t_0)=-\beta_{r-1}$ and $\psi_{r-1}(t)<-\beta_{r-1}$ on $(t_0,t_1]$. On that interval, strict monotonicity gives $\alpha_r(\psi_{r-1})<\alpha_r(-\beta_{r-1})\leq-\beta_r$, hence $\dot{\psi}_{r-1}>0$ a.e., so $\psi_{r-1}(t_1)=\psi_{r-1}(t_0)+\int_{t_0}^{t_1}\dot{\psi}_{r-1}\,dt>-\beta_{r-1}$, a contradiction. Thus $\psi_{r-1}(t)\geq-\beta_{r-1}$ on $[0,T]$. Now suppose $\psi_k(t)\geq-\beta_k$ on $[0,T]$ for some $k\in\{1,\ldots,r-1\}$. Since $\psi_k=\dot{\psi}_{k-1}+\alpha_k(\psi_{k-1})$, we have $\dot{\psi}_{k-1}(t)\geq-\beta_k-\alpha_k(\psi_{k-1}(t))$ a.e., and the same inward pointing argument with $\alpha_k(-\beta_{k-1})\leq-\beta_k$ gives $\psi_{k-1}(t)\geq-\beta_{k-1}$. Repeating down to $k=1$ proves the bound, and $\psi_0=h$ gives $h\geq-\beta_0$. When each $\alpha_k$ is onto and strictly increasing, $\alpha_k(-\beta_{k-1})=-\beta_k$ has the unique solution $\beta_{k-1}=-\alpha_k^{-1}(-\beta_k)$, the least admissible choice; for $\alpha_k(s)=\gamma_k s$ this is $\beta_{k-1}=\beta_k/\gamma_k$, and the product formula~\eqref{eq:linear_margin} follows.
\end{proof}

Theorem~\ref{thm:dae_issf_margin} gives an explicit bound from the HOCBF gains. The next corollary applies it to sampled-data command holding.

\begin{mycor}[Sampled-Data Inflation Under Command Hold]\label{cor:sampled_data_inflation}
Consider samples $0=t_0<t_1<\cdots<t_N=T$ with $t_{k+1}-t_k\leq\Delta t$, and hold the supervisory input constant on each interval. For $t\in[t_k,t_{k+1})$, define
\[
    \widetilde{\psi}_r(t):=A_r+\m{B}_r^\top\m{u}(t_k)+\underline{D}_r+\pi_{r-1}
\]
along the trajectory. If $\widetilde{\psi}_r(t_k)\geq0$ at every sample and $|\dot{\widetilde{\psi}}_r(t)|\leq L$ a.e.\ on the swept compact tube, then $\psi_r(t)\geq-L\Delta t$ a.e. Hence Theorem~\ref{thm:dae_issf_margin} applies with $\delta\equiv L\Delta t$. For linear gains, $h(\m{x}(t))\geq -L\Delta t/\prod_{k=1}^{r}\gamma_k$ on $[0,T]$.
\end{mycor}

\begin{proof}
For $t\in[t_k,t_{k+1})$, $\widetilde{\psi}_r(t)\geq\widetilde{\psi}_r(t_k)-L(t-t_k)\geq-L\Delta t$. The actual last term satisfies $\psi_r(t)\geq\widetilde{\psi}_r(t)$ because $D_r\geq\underline{D}_r$ under the same held input, so $\psi_r(t)\geq-L\Delta t$ a.e., and Theorem~\ref{thm:dae_issf_margin} gives the stated bound.
\end{proof}

Corollary~\ref{cor:sampled_data_inflation} is the formal counterpart of the one step sampled-data gap reported in the implementation (Section~\ref{subsec:impl}): the safe set inflation is $O(\Delta t)$ and shrinks as the control rate increases. For power system applications the natural initial condition is an operating point before the disturbance, which automatically satisfies the nested initialization of Theorem~\ref{thm:forward_inv}, as the next lemma records.

\begin{mylem}[Initial Condition Compatibility at Equilibrium]\label{lem:equil_compat}
Let $\m{x}^\star$ be an equilibrium of~\eqref{eq:std_form} under constant $\m{w} = \m{w}^\star$ with $h(\m{x}^\star)\geq 0$. Then $\psi_k(0)\geq 0$ holds at $(\m{x}^\star,(\m{w}^\star,\m{0},\ldots,\m{0}))$ for all $k=0,\ldots,r-1$.
\end{mylem}

\begin{proof}
At equilibrium under constant disturbance the trajectory is constant, $\m{x}(t)\equiv \m{x}^\star$ and $\m{w}(t)\equiv \m{w}^\star$, so every time derivative of every smooth trajectory functional vanishes, in particular $\dot\psi_{k-1}(0)=0$ for each $k\geq 1$. By induction, $\psi_0(0)=h(\m{x}^\star)\geq 0$, and if $\psi_{k-1}(0)\geq 0$ then $\psi_k(0) = \dot\psi_{k-1}(0) + \alpha_k(\psi_{k-1}(0)) = \alpha_k(\psi_{k-1}(0)) \geq 0$.
\end{proof}

The initial condition hypotheses of Theorem~\ref{thm:forward_inv} therefore hold at any equilibrium before the disturbance on $\mathcal{C}_0\cap\mathcal{D}$, and by continuity of the $\psi_k$ on an open neighborhood thereof. The remaining hypothesis, that the trajectory stays in $\mathcal{D}$ on $[0,T]$, is a standard validity domain requirement for local nonlinear control results and is checked a posteriori by the reachability analysis of Section~\ref{sec:CORA} applied to the closed-loop system. Theorem~\ref{thm:forward_inv} should thus be read as an invariance result conditional on the domain: combined with a separate verification that the reachable set stays inside $\mathcal{D}$, it yields the hard safety guarantee on $[0,T]$. Cases in which the state lies outside $\mathcal{C}_\cap$ when the filter engages are recovery problems and lie outside the scope of the present theorem.

\begin{myrem}[Compatibility Through Local Manifold Reduction]\label{rem:manifold_auto}
In the general DAE-CBF framework~\cite{zhang2026verification}, compatibility between a barrier condition and the algebraic constraint manifold requires explicit geometric conditions. In the present semi-explicit index-1 setting, Assumption~\ref{assump:main}\textit{(i)} yields the local reduction $\m{x}_a=\m{\varphi}(\m{x}_d,\m{w})$, so the barrier dynamics can be expressed on the constraint manifold through the reduced ODE~\eqref{eq:reduced_ODE} and the projected gradient~\eqref{eq:proj_grad}. This local reduction is the mechanism used here to avoid introducing derivatives of the supervisory input. No broader automatic compatibility claim is made beyond the validity domain of that reduction.
\end{myrem}

\begin{myrem}[Multi-Input Coupling]\label{rem:multi_input}
When $n_u > 1$ and $\m{B}_r$ has nonzero entries in several components of $\m{u}$, the HOCBF constraint couples those components, and when multiple barriers are combined in the QP, cross-barrier coupling arises whenever their $\m{B}_{r_j}$ vectors share components. The feasibility analysis of~\cite{xiao2022control} applies directly.
\end{myrem}

\subsection{QP Safety Filter and Closed Form Control Law}\label{subsec:QP}

With the invariance result in hand, the online filter enforces the DAE-HOCBF inequalities for all safety outputs at each time instant. For a family of barriers $\{h_j\}_{j=1}^{N_h}$ with relative degrees $\{r_j\}$ on domains $\{\mathcal{D}_j\}$, denote by $A_{r_j,j}, \m{B}_{r_j,j}, \underline{D}_{r_j,j}, \pi_{r_j-1,j}$ the quantities built from $h_j$ via~\eqref{eq:psi_seq}--\eqref{eq:HOCBF_cond}, and set
\begin{equation}\label{eq:cj}
    c_j(\m{x},\m{w}^{[r_j-1]}) := -\underline{D}_{r_j,j}(\m{x},\m{w}^{[r_j-1]}) - \pi_{r_j-1,j}(\m{x},\m{w}^{[r_j-1]}).
\end{equation}
The filter solves
\begin{subequations}\label{eq:QP}
\begin{align}
    (\m{u}^*,\m{\xi}^*) = \arg\min_{\m{u},\m{\xi}} \;\; & \|\m{u}-\m{u}_{\text{nom}}\|^2 + \kappa\|\m{\xi}\|^2 \label{eq:QP_obj}\\
    \text{s.t.} \;\; & A_{r_j,j} + \m{B}_{r_j,j}^\top\m{u} \geq c_j - \xi_j, \quad j=1,\ldots,N_h, \label{eq:QP_rdj}\\
    & \m{u}_{\min}\leq \m{u}\leq \m{u}_{\max}, \; \m{\xi}\geq \m{0}. \label{eq:QP_bounds}
\end{align}
\end{subequations}
Problem~\eqref{eq:QP} returns a supervisory command close to the nominal output while penalizing violation of the DAE-HOCBF rows. The slack $\m{\xi}\geq\m{0}$ keeps the problem feasible when the hard rows and actuator box are incompatible. Hard safety follows when $\m{\xi}^*=\m{0}$ and the hypotheses of Theorem~\ref{thm:forward_inv} hold. If $\m{\xi}^*\neq\m{0}$, Corollary~\ref{cor:slack_inflation} bounds the induced safe set inflation. The QP is static in $\m{u}$ because Corollary~\ref{cor:chain_rule} stops the expansion at each relative degree $r_j$, before any derivative of $\m{u}$ appears.

\begin{mycor}[Slack Induced Safety Inflation]\label{cor:slack_inflation}
For barrier $h_j$ in the soft QP~\eqref{eq:QP}, let $\bar{\xi}_j\geq0$ satisfy $\xi_j^*(t)\leq\bar{\xi}_j$ for a.e.\ $t\in[0,T]$. If the trajectory remains in $\mathcal{D}_j$ and the initial condition satisfies the inflated bounds of Theorem~\ref{thm:dae_issf_margin}, then $h_j(\m{x}(t))\geq-\beta_{0,j}$ on $[0,T]$, with $\beta_{0,j}$ computed from that theorem using $\delta=\xi_j^*$. For linear gains, $\beta_{0,j}=\bar{\xi}_j/\prod_{k=1}^{r_j}\gamma_{k,j}$.
\end{mycor}

\begin{proof}
The $j$th QP row gives $A_{r_j,j}+\m{B}_{r_j,j}^{\top}\m{u}^*+\underline{D}_{r_j,j}+\pi_{r_j-1,j}\geq -\xi_j^*$. Along the actual trajectory $D_{r_j,j}\geq\underline{D}_{r_j,j}$ by definition of the worst case tightening, so $\psi_{r_j,j}=A_{r_j,j}+\m{B}_{r_j,j}^{\top}\m{u}^*+D_{r_j,j}+\pi_{r_j-1,j}\geq -\xi_j^*$. Applying Theorem~\ref{thm:dae_issf_margin} with $\delta=\xi_j^*$ gives the result.
\end{proof}

We first record the closed form solution in the simplest regime, then establish the regularity of the general filter.

\vspace{0.1cm}
\noindent\textbf{Closed form law in the single-row regime.} When one barrier row is relevant and actuator bounds and other barriers are inactive, the soft QP can be solved in closed form. The formula is useful as a sanity check: finite $\kappa$ splits a nominal violation between control motion and slack, while the hard projection is recovered as $\kappa\to\infty$.

\begin{myprs}[Closed Form Single-Row Soft Filter]\label{prop:closed_form}
Consider one row $a+\m{b}^\top\m{u}\geq c-\xi$ with $\xi\geq0$, no active box constraint, and penalty $\kappa>0$. Let $d:=c-a-\m{b}^\top\m{u}_{\emph{nom}}$. The unique minimizer of
\[
    \min_{\m{u},\xi\geq0}\;\|\m{u}-\m{u}_{\emph{nom}}\|^2+\kappa\xi^2
    \quad\text{s.t.}\quad a+\m{b}^\top\m{u}\geq c-\xi
\]
is
\begin{equation}\label{eq:closed_form}
    \m{u}^*=\m{u}_{\emph{nom}}+\lambda^*\m{b},\qquad
    \xi^*=\frac{\lambda^*}{\kappa},\qquad
    \lambda^*=\frac{(d)_+}{\|\m{b}\|^2+\kappa^{-1}}.
\end{equation}
If $a$, $\m{b}$, $c$, and $\m{u}_{\emph{nom}}$ are locally Lipschitz in the state variables, then so are $\m{u}^*$ and $\xi^*$. As $\kappa\to\infty$ and $\m{b}\neq0$, $\m{u}^*$ converges to the hard projection onto the half space, $\m{u}_{\emph{nom}}+(d)_+\m{b}/\|\m{b}\|^2$, and $\xi^*\to0$.
\end{myprs}

\begin{proof}
If $d\leq0$, the nominal input satisfies the row and the minimizer is $(\m{u}_{\emph{nom}},0)$. If $d>0$, write $\Delta\m{u}=\m{u}-\m{u}_{\emph{nom}}$. The active constraint is $\m{b}^\top\Delta\m{u}+\xi=d$. The KKT conditions for the strictly convex problem give $2\Delta\m{u}=\mu\m{b}$ and $2\kappa\xi=\mu$. Substitution into the active constraint yields $\mu=2d/(\|\m{b}\|^2+\kappa^{-1})$, which gives~\eqref{eq:closed_form}. Local Lipschitz continuity follows because $(\cdot)_+$ is Lipschitz and the denominator is bounded below by $\kappa^{-1}$.
\end{proof}

Beyond this regime the filter is solved as the full QP~\eqref{eq:QP}. The next proposition records that the DAE-HOCBF construction does not change the online optimization class.

\begin{myprs}[QP Class Preservation]\label{prop:qp_class}
At each time instant, if $\kappa>0$ and the actuator box is nonempty, problem~\eqref{eq:QP} is a feasible strictly convex QP in $(\m{u},\m{\xi})\in\mathbb{R}^{n_u+N_h}$. It has $2N_h+2n_u$ linear inequality constraints and a unique minimizer. The disturbance tightening and the multiple barrier rows only change affine coefficients.
\end{myprs}

\begin{proof}
The objective has Hessian $\mathrm{blkdiag}(2\m{I}_{n_u},2\kappa\m{I}_{N_h})\succ0$, so it is strictly convex. For fixed current data, Lemma~\ref{lem:dae_lift} makes $A_{r_j,j}$ and $\m{B}_{r_j,j}$ fixed coefficients, while $\underline{D}_{r_j,j}$ and $\pi_{r_j-1,j}$ evaluate to scalars that shift the right-hand side of the row~\eqref{eq:QP_rdj}. Multiple barriers stack such affine rows in the shared variable $\m{u}$, and the bounds and slack nonnegativity in~\eqref{eq:QP_bounds} are linear. Feasibility follows from the nonempty actuator box and sufficiently large slacks, so the feasible set is a nonempty polyhedron and the strictly convex QP has a unique minimizer.
\end{proof}

The closed-loop and reachability construction also needs regular dependence of the optimizer on the current data. The soft formulation provides this without an active set or linear independence assumption, because the slack can be eliminated and the remaining feasible set is the fixed actuator box. We first state the parametric optimization fact used for this step.

\begin{mylem}[Fixed Set Strongly Convex Minimizer]\label{lem:fixed_set_sc_lip}
Let $\mathcal{U}\subset\mathbb{R}^{n_u}$ be closed and convex. Suppose $F:\Theta\times\mathcal{U}\to\mathbb{R}$ is continuously differentiable in $\m{u}$, uniformly $m$-strongly convex in $\m{u}$, and has a gradient that is Lipschitz in the parameter:
\[
    \|\nabla_{\m{u}}F(\m{\theta}_1,\m{u})-\nabla_{\m{u}}F(\m{\theta}_2,\m{u})\|\leq L_{\theta}\|\m{\theta}_1-\m{\theta}_2\|.
\]
Then $\m{u}^*(\m{\theta}):=\arg\min_{\m{u}\in\mathcal{U}}F(\m{\theta},\m{u})$ is single-valued and Lipschitz, with constant at most $L_{\theta}/m$.
\end{mylem}

\begin{proof}
Strong convexity gives uniqueness. Let $\m{u}_i=\m{u}^*(\m{\theta}_i)$ and $\m{d}:=\m{u}_1-\m{u}_2$. The variational inequalities over the fixed convex set $\mathcal{U}$ give $\nabla_{\m{u}}F(\m{\theta}_1,\m{u}_1)^\top(\m{u}_2-\m{u}_1)\geq 0$ and $\nabla_{\m{u}}F(\m{\theta}_2,\m{u}_2)^\top(\m{u}_1-\m{u}_2)\geq 0$, which add to $(\nabla_{\m{u}}F(\m{\theta}_1,\m{u}_1)-\nabla_{\m{u}}F(\m{\theta}_2,\m{u}_2))^\top\m{d}\leq 0$. Then
\[
    \begin{aligned}
    m\|\m{d}\|^2
    &\leq \bigl(\nabla_{\m{u}}F(\m{\theta}_1,\m{u}_1)
    -\nabla_{\m{u}}F(\m{\theta}_1,\m{u}_2)\bigr)^\top\m{d}\\
    &\leq \bigl(\nabla_{\m{u}}F(\m{\theta}_2,\m{u}_2)
    -\nabla_{\m{u}}F(\m{\theta}_1,\m{u}_2)\bigr)^\top\m{d}\\
    &\leq L_{\theta}\|\m{\theta}_1-\m{\theta}_2\|\,\|\m{d}\|,
    \end{aligned}
\]
where the first inequality is $m$-strong convexity, the second uses $(\nabla_{\m{u}}F(\m{\theta}_1,\m{u}_1)-\nabla_{\m{u}}F(\m{\theta}_2,\m{u}_2))^\top\m{d}\leq0$, and the third is the Lipschitz bound with Cauchy-Schwarz. Dividing by $\|\m{d}\|$ when $\m{d}\neq\m{0}$ gives the result, and the bound is trivial otherwise.
\end{proof}

\begin{theorem}[Locally Lipschitz Soft QP Safety Filter]\label{thm:soft_qp_lipschitz}
Consider the soft QP~\eqref{eq:QP} with fixed nonempty actuator box and $\kappa>0$. Collect its current data as $\m{\theta}:=(\m{u}_{\emph{nom}},\{A_{r_j,j},\m{B}_{r_j,j},c_j\}_{j=1}^{N_h})$. On any compact set $\Theta$ where these data are bounded and Lipschitz, the QP control law $\m{u}^*(\m{\theta})$ is Lipschitz.

Equivalently, after eliminating the optimal slack,
\begin{equation}\label{eq:reduced_soft_qp}
    F_{\kappa}(\m{\theta},\m{u}) := \|\m{u}-\m{u}_{\emph{nom}}\|^2 +\kappa\sum_{j=1}^{N_h}\bigl(c_j-A_{r_j,j}-\m{B}_{r_j,j}^\top\m{u}\bigr)_+^2 ,
\end{equation}
and $\m{u}^*(\m{\theta})$ is the unique minimizer of $F_\kappa$ over $\mathcal{U}$. No active set constancy, LICQ, or restriction to a single row is needed.
\end{theorem}

\begin{proof}
For fixed $\m{\theta}$ and $\m{u}$ the constraints on $\xi_j$ are $\xi_j\geq c_j-A_{r_j,j}-\m{B}_{r_j,j}^{\top}\m{u}$ and $\xi_j\geq 0$, and since the objective increases in $\xi_j^2$ the optimal slack is $\xi_j=(c_j-A_{r_j,j}-\m{B}_{r_j,j}^{\top}\m{u})_+$. Substituting it reduces the joint problem, in its control component, to minimizing~\eqref{eq:reduced_soft_qp} over the fixed box $\mathcal{U}$. The map $s\mapsto(s)_+^2$ is $C^1$ and convex, and each row residual is affine in $\m{u}$, so the penalty sum is convex and $F_{\kappa}$ is uniformly $2$-strongly convex in $\m{u}$. Its gradient $\nabla_{\m{u}}F_{\kappa}=2(\m{u}-\m{u}_{\emph{nom}})-2\kappa\sum_j(c_j-A_{r_j,j}-\m{B}_{r_j,j}^{\top}\m{u})_+\m{B}_{r_j,j}$ is, on the compact set $\Theta$, Lipschitz in $\m{\theta}$ uniformly over $\m{u}\in\mathcal{U}$, because the data are bounded and Lipschitz and $(\cdot)_+$ is $1$-Lipschitz. Lemma~\ref{lem:fixed_set_sc_lip} with $m=2$ then gives Lipschitz continuity of $\m{u}^*(\m{\theta})$.
\end{proof}

The decomposition in~\eqref{eq:reduced_soft_qp} also shows the online cost splits into coefficient assembly plus a QP solve. The former is system-specific and is assessed empirically in Section~\ref{subsec:impl}, while the latter is handled by standard real-time QP solvers and certified using existing active set complexity results~\cite{arnstrom2021unifying,arnstrom2024exact}. Since $\m{w}^{[r-1]}$ is an exogenous bounded signal by Assumption~\ref{assump:main}\textit{(iii)}, the relevant regularity for Section~\ref{sec:CORA} is local Lipschitz continuity of $\m{u}^*$ in $\m{x}$ at fixed $\m{w}^{[r-1]}$, which Theorem~\ref{thm:soft_qp_lipschitz} supplies. In the single-row, inactive-box closed form regime the filter splits a nominal violation between a correction along $\m{b}$ and slack, and returns $\m{u}^*=\m{u}_{\text{nom}}$ with zero slack whenever the nominal controller is already safe.

Algorithm~\ref{alg:framework} summarizes the resulting pipeline. It instantiates the DAE-HOCBF construction for the chosen implemented plant and controller stack, runs the online safety filtering loop, and states the offline reachability step developed in Section~\ref{sec:CORA}.

\begin{algorithm}[t]
\caption{DAE-HOCBF Filter-and-Verify Pipeline}
\label{alg:framework}
\small
\begin{algorithmic}[1]
\Require DAE model~\eqref{eq:DAE}; supervisory input channels; nominal controller $\m{u}_{\emph{nom}}$; safety functions $\{h_j\}_{j=1}^{N_h}$ defining~\eqref{eq:safe_set}; actuator bounds $\mathcal{U}$; disturbance class $\mathcal{W}_{r_{\max}}$; algebraically consistent initial set $\mathcal{X}_0$; horizon $[0,T]$
\Ensure Safety filter $\mathcal{F}$ and verification result on $[0,T]$

\State Fix the implemented plant and controller stack.
\If{$\partial \m{g}/\partial \m{u} \not\equiv \m{0}$}
    \State Augment with the pre-filter~\eqref{eq:prefilter} to obtain~\eqref{eq:aug_DAE}; set the new command as the supervisory input.
\EndIf
\State Express the implemented plant in standard form~\eqref{eq:std_form}.
\For{$j=1,\dots,N_h$}
    \State Determine the relative degree $r_j$ of $h_j$.
    \State Construct $\psi_{0,j},\dots,\psi_{r_j,j}$ via~\eqref{eq:psi_seq}.
    \State Compute the HOCBF data $A_{r_j,j}, \m{B}_{r_j,j}, \underline{D}_{r_j,j}, \pi_{r_j-1,j}$ in~\eqref{eq:HOCBF_cond}.
    \State Evaluate the residual~\eqref{eq:rho_h_def}; by Theorem~\ref{thm:robust_row_feasibility}, nonemptiness of~\eqref{eq:adm_set} is equivalent to $\rho_{h_j}\geq0$.
\EndFor

\Statex \textbf{Online filtering}
\While{the system is in operation}
    \State Measure or estimate $(\m{x},\m{w}^{[r_{\max}-1]})$.
    \State Evaluate all DAE-HOCBF constraints~\eqref{eq:HOCBF_cond}.
    \State Solve~\eqref{eq:QP} for $(\m{u}^*,\m{\xi}^*)$; use~\eqref{eq:closed_form} when the single-row, inactive-box conditions hold.
    \State Apply $\m{u}^*$ to the implemented plant.
\EndWhile

\Statex \textbf{Offline verification}
\State Encode disturbance derivatives by the integrator chain exosystem.
\State Form the closed-loop DAE with augmented state by interconnecting the nominal controller, safety filter, plant, and exosystem.
\State Compute a sound reachable set enclosure on $[0,T]$ for all $\m{x}(0)\in\mathcal{X}_0$ and all admissible disturbances.
\State Declare safety if the reachable set does not intersect the unsafe set; otherwise refine the setup and repeat.
\end{algorithmic}
\end{algorithm}

\section{Reachability Verification of the Closed-Loop DAE}\label{sec:CORA}

We now turn from online enforcement to offline certification of the closed-loop system. The HOCBF row depends on $\m{x}$ and on $\m{w}^{[r-1]}$, so reachability must be run on an augmented system whose state contains the needed disturbance derivatives. This section builds that system. It applies directly to~\eqref{eq:DAE} when $\partial\m{g}/\partial \m{u}\equiv\m{0}$ and to the pre-filtered plant~\eqref{eq:aug_DAE} otherwise.

\subsection{Disturbance Exosystem}\label{subsec:exosystem}

The first step is to encode the disturbance derivatives as exosystem states. Let $r_{\max}$ be the largest relative degree among the barriers, and introduce
\begin{equation}\label{eq:exosystem}
    \dot{\m{\eta}}_\ell = \m{\eta}_{\ell+1}, \;\; \ell = 0,\ldots,r_{\max}-1, \qquad \m{\eta}_{r_{\max}}(t)\in\mathcal{V} := \{\m{v}:\|\m{v}\|\leq \bar{w}_{r_{\max}}\}.
\end{equation}
We identify $\m{\eta}_\ell$ with $\m{w}^{(\ell)}$ for $\ell=0,\ldots,r_{\max}-1$ and treat $\m{\eta}_{r_{\max}}$ as the bounded exogenous input. Initial conditions are taken over the admissible derivative sets $\m{\eta}_0(0)\in\mathcal{W}$ and $\m{\eta}_\ell(0)\in\mathcal{E}_\ell$ for $\ell=1,\ldots,r_{\max}-1$, where
\[
    \mathcal{E}_\ell:=\{\m{v}\in\mathbb{R}^{n_w}:\|\m{v}\|\leq\bar{w}_\ell\}.
\]
Together with $\mathcal{V}$ for the top derivative input, this exosystem overbounds all disturbances in $\mathcal{W}_{r_{\max}}$.

\subsection{Closed-Loop Verified System}\label{subsec:closed_loop}

With the exosystem in place, we couple the plant~\eqref{eq:std_form}, the QP safety filter~\eqref{eq:QP}, and the exosystem~\eqref{eq:exosystem}. The nominal controller is represented in either of the following standard forms:
\begin{enumerate}[label=(N\arabic*)]
    \item \emph{Static feedback}, a locally Lipschitz state feedback map $\m{u}_{\text{nom}} = \m{\kappa}(\m{x}_d,\m{x}_a,\m{\eta}_0)$, such as AVR/governor droop laws or learned feedback policies;
    \item \emph{Dynamic feedback with internal states}, with smooth dynamics $\dot{\m{x}}_c = \m{f}_c(\m{x}_c,\m{x}_d,\m{x}_a,\m{\eta}_0)$ and output $\m{u}_{\text{nom}} = \m{\kappa}(\m{x}_c,\m{x}_d,\m{x}_a,\m{\eta}_0)$, such as integral or PSS dynamics, whose internal states $\m{x}_c$ are absorbed into the dynamic state by redefining $\m{x}_d \leftarrow [\m{x}_d^\top,\m{x}_c^\top]^\top$ prior to the construction of Section~\ref{sec:CBF}.
\end{enumerate}
Under either form $\m{u}_{\text{nom}}$ is a locally Lipschitz function of the augmented state. On every compact subset of the validity domain where the HOCBF coefficient maps are bounded and Lipschitz, Theorem~\ref{thm:soft_qp_lipschitz} makes the soft QP map $\m{u}^*$ locally Lipschitz in the augmented state and algebraic variable, which supplies the regularity hypothesis used below. The verified system has augmented state
\begin{equation}\label{eq:aug_state}
    \m{z} := \bigl[\m{x}_d^\top,\;\m{\eta}_0^\top,\m{\eta}_1^\top,\ldots,\m{\eta}_{r_{\max}-1}^\top\bigr]^\top
\end{equation}
and augmented algebraic variable $\m{y} := \m{x}_a$. Writing the QP output as $\m{u}^*(\m{z},\m{y})$ and identifying $\m{w} = \m{\eta}_0$, $\m{w}^{(\ell)} = \m{\eta}_\ell$, the closed-loop DAE reads
\begin{subequations}\label{eq:CL_DAE}
\begin{align}
    \dot{\m{x}}_d &= \m{f}_0(\m{x}_d, \m{y}, \m{\eta}_0) + \m{B}_d(\m{x}_d, \m{y}, \m{\eta}_0)\,\m{u}^*(\m{z},\m{y}), \label{eq:CL_DAE_a}\\
    \dot{\m{\eta}}_\ell &= \m{\eta}_{\ell+1},\;\; \ell = 0,\ldots,r_{\max}-2, \label{eq:CL_DAE_b}\\
    \dot{\m{\eta}}_{r_{\max}-1} &= \m{v}(t), \qquad \m{v}(t)\in\mathcal{V}, \label{eq:CL_DAE_c}\\
    \m{0} &= \m{g}(\m{x}_d, \m{y}, \m{\eta}_0). \label{eq:CL_DAE_d}
\end{align}
\end{subequations}
System~\eqref{eq:CL_DAE} is the object verified offline. It is again a semi-explicit index-1 DAE, now driven by the bounded input $\m{v}\in\mathcal{V}$.

\begin{myprs}[Closed-Loop DAE Is Well-Posed]\label{prop:CL_wellposed}
If the QP map $\m{u}^*(\m{z},\m{y})$ is locally Lipschitz in $(\m{z},\m{y})$, then~\eqref{eq:CL_DAE} admits a locally unique absolutely continuous solution for every $\m{v}\in L^\infty([0,T];\mathcal{V})$ and every algebraically consistent initial state.
\end{myprs}

\begin{proof}
The algebraic Jacobian of~\eqref{eq:CL_DAE_d} with respect to $\m{y}$ is $\m{J}_a$, nonsingular by Assumption~\ref{assump:main}\textit{(i)}, so locally around any algebraically consistent point the implicit function theorem gives a $C^{s-1}$ map $\m{y}=\m{\varphi}_{\mathrm{cl}}(\m{z})$ parametrizing the consistency manifold. Substituting it into~\eqref{eq:CL_DAE_a}--\eqref{eq:CL_DAE_c} gives a reduced ODE $\dot{\m{z}} = \bar{\m{F}}(\m{z},\m{v}(t))$ whose components compose the $C^s$ maps $\m{f}_0,\m{B}_d$ with the $C^{s-1}$ map $\m{\varphi}_{\mathrm{cl}}$ and the locally Lipschitz map $\m{u}^*(\m{z},\m{\varphi}_{\mathrm{cl}}(\m{z}))$. Hence $\bar{\m{F}}(\cdot,\m{v})$ is locally Lipschitz in $\m{z}$ uniformly in $\m{v}\in\mathcal{V}$, and $t\mapsto \bar{\m{F}}(\m{z},\m{v}(t))$ is measurable for any $\m{v}\in L^\infty([0,T];\mathcal{V})$. Carathéodory existence and uniqueness give a locally unique absolutely continuous $\m{z}(\cdot)$~\cite[Thm.~3.1]{khalil2002nonlinear}, and $\m{y}(t):=\m{\varphi}_{\mathrm{cl}}(\m{z}(t))$ is absolutely continuous because $\m{\varphi}_{\mathrm{cl}}$ is $C^{s-1}$, so~\eqref{eq:CL_DAE} has a locally unique absolutely continuous algebraically consistent solution.
\end{proof}

The Lipschitz hypothesis holds in the single-row, inactive-box regime by Proposition~\ref{prop:closed_form}. It also holds for the soft filter with multiple constraints by Theorem~\ref{thm:soft_qp_lipschitz} on compact subsets where the HOCBF coefficient maps are bounded and Lipschitz.

\subsection{Verification Condition}\label{subsec:verification}

Once the augmented model is fixed, verification reduces to a separation test between the reachable set and the unsafe set. Reachability analysis computes an overapproximation $\hat{\mathcal{R}}([0,T])$ of the forward reachable set of~\eqref{eq:CL_DAE} on $[0,T]$ from initial conditions
\begin{equation}\label{eq:IC_set}
    (\m{x}_d(0),\m{y}(0))\in\mathcal{X}_0\subset\mathcal{C}, \;\; \m{\eta}_0(0)\in\mathcal{W}, \;\; \m{\eta}_\ell(0)\in\mathcal{E}_\ell,\;\ell=1,\ldots,r_{\max}-1,
\end{equation}
under all admissible $\m{v}\in L^\infty([0,T];\mathcal{V})$, with unsafe set
\begin{equation}\label{eq:unsafe}
    \mathcal{U}_{\text{unsafe}} := \{(\m{x}_d,\m{y}):(\m{x}_d,\m{y})\notin\mathcal{C}\} = \bigcup_{j=1}^{N_h}\{h_j < 0\}.
\end{equation}

\begin{mydef}[Verified Safety Certificate]\label{def:cert}
The closed-loop DAE~\eqref{eq:CL_DAE} is \emph{verified safe} on $[0,T]$ w.r.t.\ Definition~\ref{def:safety} if $\Pi_{(\m{x}_d,\m{y})}\hat{\mathcal{R}}([0,T]) \cap \mathcal{U}_{\emph{unsafe}} = \emptyset$, where $\Pi_{(\m{x}_d,\m{y})}$ projects out the exosystem variables.
\end{mydef}

This certificate is sound and one-sided. If the overapproximation misses the unsafe set, then the true reachable set is safe. The formulation~\eqref{eq:CL_DAE} can be passed to any sound nonlinear semi-explicit index-1 DAE reachability method. Section~\ref{sec:case_studies} uses a zonotope engine specialized to this structure.

\section{Case Studies}\label{sec:case_studies}
This section instantiates the framework on two standard power system benchmarks, covering setup, implementation, time domain studies, reachability certification with Section~\ref{sec:CORA}, and limitations.

\subsection{Test Systems, Controller Stacks, and Scenarios}\label{subsec:testsys}

All simulations use ANDES~\cite{cui2020hybrid}, an open source Python toolbox whose device library represents generators, exciters, governors, and stabilizers as hybrid DAEs with explicit dynamic and algebraic equations rather than reduced-order ODE surrogates. We use two ANDES benchmark networks. Both have the same round-rotor machine model (GENROU), but different exciter and governor stacks, which changes the relative degree of every supervised barrier. Table~\ref{tab:case_study_setup} summarizes the supervision map and Table~\ref{tab:relative_degree_audit} reports the neighborhood audit.

\begin{table}[t]
\centering
\setlength{\tabcolsep}{3.5pt}
\renewcommand{\arraystretch}{1.15}
\caption{Case study setup and supervision map for the implemented ANDES DAE-HOCBF experiments.}
\label{tab:case_study_setup}
\resizebox{\columnwidth}{!}{%
\begin{tabular}{lllccccc}
\toprule
\multicolumn{1}{c}{Case} &
\multicolumn{1}{c}{Scenario} &
\multicolumn{1}{c}{\makecell[c]{Supervised\\quantity}} &
\multicolumn{1}{c}{Barrier form} &
\multicolumn{1}{c}{\makecell[c]{Supervisory\\channel}} &
\multicolumn{1}{c}{\makecell[c]{Relative\\degree $r$}} &
\multicolumn{1}{c}{$n_{\text{barriers}}$} &
\multicolumn{1}{c}{$n_u$} \\
\midrule
\multicolumn{8}{l}{\textbf{Kundur}} \\
Kundur & Load ramp & Voltage & $(v-v_{\min})(v_{\max}-v)$ & $v_{\mathrm{ref}}$ & $4$ & $4$ & $4$ \\
Kundur & Generator trip & Frequency & $\Delta\omega_{\max}^2-(\omega-1)^2$ & $p_{\mathrm{aux}}$ & $3$ & $4$ & $4$ \\
Kundur & Load ramp & Voltage + freq & $h_v,\;h_\omega$ & $v_{\mathrm{ref}}$, $p_{\mathrm{aux}}$ & $r_v=4,\;r_\omega=3$ & $8$ & $8$ \\
\midrule
\multicolumn{8}{l}{\textbf{IEEE-39}} \\
IEEE-39 & Generator trip & Frequency & $\Delta\omega_{\max}^2-(\omega-1)^2$ & $p_{\mathrm{aux}}$ & $3$ & $10$ & $10$ \\
IEEE-39 & Voltage moderate & Voltage & $(v-v_{\min})(v_{\max}-v)$ & $v_{\mathrm{ref}}$ & $5$ & $7$ & $10$ \\
IEEE-39 & Voltage severe & Voltage & $(v-v_{\min})(v_{\max}-v)$ & $v_{\mathrm{ref}}$ & $5$ & $7$ & $10$ \\
\bottomrule
\end{tabular}
}
\end{table}

\begin{table}[t]
\vspace{0.2cm}
\centering
\setlength{\tabcolsep}{3.5pt}
\renewcommand{\arraystretch}{1.15}
\caption{Neighborhood relative degree audit. The observed relative degrees are stable over sampled perturbations around the nominal operating point. The reference threshold is $\mathrm{tol}=10^{-6}$; ratios above unity certify non-vanishing of the control direction Lie derivative.}
\label{tab:relative_degree_audit}
\resizebox{\columnwidth}{!}{%
\begin{tabular}{llcccc}
\toprule
Case & Barrier family & $r$ & $n_{\text{samples}}$ & min $\left|L_{\bar b}L_{\bar F}^{r-1}\bar H\right|$ & min ratio $\left|L_{\bar b}L_{\bar F}^{r-1}\bar H\right| / \text{tol}$ \\
\midrule
\multicolumn{6}{l}{\textbf{Kundur}} \\
Kundur voltage & Voltage & $4$ & $30$ & $4.372$ & $4.37\!\times\!10^{6}$ \\
Kundur frequency & Frequency & $3$ & $30$ & $5.57\!\times\!10^{-3}$ & $5.57\!\times\!10^{3}$ \\
Kundur combined (voltage) & Voltage & $4$ & $30$ & $1.958$ & $1.96\!\times\!10^{6}$ \\
Kundur combined (frequency) & Frequency & $3$ & $30$ & $0.012$ & $1.18\!\times\!10^{4}$ \\
\midrule
\multicolumn{6}{l}{\textbf{IEEE-39}} \\
IEEE-39 frequency & Frequency & $3$ & $30$ & $2.85\!\times\!10^{-4}$ & $2.85\!\times\!10^{2}$ \\
IEEE-39 voltage moderate & Voltage & $5$ & $30$ & $1.016$ & $1.02\!\times\!10^{6}$ \\
IEEE-39 voltage severe & Voltage & $5$ & $30$ & $1.310$ & $1.31\!\times\!10^{6}$ \\
\bottomrule
\end{tabular}
}
\end{table}

\vspace{0.1cm}
\noindent\textbf{Kundur two-area system.} The Kundur case has four generators with EXDC2 exciters and TGOV1 governors. We supervise terminal voltages $v_1,\ldots,v_4$ through exciter $v_{\mathrm{ref}}$ channels and generator speeds $\omega_1,\ldots,\omega_4$ through governor $p_{\mathrm{aux}}$ channels. With the pre-filter, voltage barriers have relative degree $4$ w.r.t.\ $v_{\mathrm{ref}}$, namely the pre-filter state, EXDC2 lag-lead internal states, and the algebraic bus voltage constraint; frequency barriers have relative degree $3$ through the pre-filter state, TGOV1 internal state, and machine inertia. The audit confirms both degrees, with minimum ratios $4.37\times 10^{6}$ and $5.57\times 10^{3}$ across the $30$-sample set.

\vspace{0.1cm}
\noindent\textbf{IEEE-39 New England system.} The IEEE-39 case has ten generators with IEEEX1 exciters and TGOV1N governors. Because IEEEX1 adds a lag state in the voltage-error path relative to EXDC2, the pre-filtered voltage barriers have relative degree $5$ w.r.t.\ $v_{\mathrm{ref}}$, while frequency barriers remain at relative degree $3$. Two generator buses in the stock case sit outside the $[0.95,1.05]$ band at the power flow solution, so voltage is supervised only on the feasible subset $\{30,31,32,33,34,37,39\}$; all ten frequency barriers are supervised. The voltage audit gives minimum ratios of $10^6$ to $10^8$, and the frequency audit ratios range from $2.85\times 10^{2}$ (gen 10) to $1.01\times 10^{4}$ (gen 7).

\vspace{0.1cm}
\noindent\textbf{Scenarios.} We use two disturbance families. A \emph{load ramp} increases every PQ load by a fraction $\alpha$ of nominal through a ninth-degree smoothstep whose first four derivatives vanish at both endpoints, which matches the disturbance class $\mathcal{W}_{r_{\max}}$ of Assumption~\ref{assump:main}\textit{(iii)} up to $r_{\max}=5$. A \emph{generator trip} disables the first generator at $t=0.5$~s and is the only hybrid event we exercise. The Kundur voltage and frequency cases use $\alpha = 0.05$ over a $0.2$~s window starting at $t=0.3$~s, with horizon $[0,1.5]$~s. The Kundur combined case uses the same ramp with tighter bounds $|\nu_v|\leq 0.05$, $|\nu_\omega|\leq 0.03$ to reflect the reduced authority when both channel families share the actuation budget. The IEEE-39 voltage cases use two stress presets, a moderate one with $\alpha=0.15$ and a $1.0$~s ramp and a severe one with $\alpha=0.20$ and a $2.0$~s ramp, with horizons $[0,2.5]$~s and $[0,3.8]$~s. For the reachability studies of Section~\ref{subsec:reach_impl} we add a \emph{benign} preset, $\alpha=0.02$ over $0.4$~s on Kundur and $0.8$~s on IEEE-39, with the same smoothstep regularity; this preset operates strictly inside the hard safety regime where the admissible input set is nonempty, the regime in which Theorem~\ref{thm:forward_inv} applies. All runs use a TDS step of $\Delta t = 0.02$~s, and the default supervisory bounds are symmetric, $|\nu_v|\leq 0.10$ and $|\nu_\omega|\leq 0.05$.

\subsection{Implementation Details and Theory vs. Practice}\label{subsec:impl}
The framework is implemented as a supervisory wrapper around ANDES, and the closed-loop DAE being filtered is the full plant assembled from device models rather than a hand derived reduced surrogate.

\vspace{0.1cm}
\noindent\textbf{Pre-filter realization.} Every supervised reference channel ($v_{\mathrm{ref}}$ for exciters, $p_{\mathrm{aux}}$ for governors) enters the ANDES controller blocks through algebraic services, so $\partial\m{g}/\partial\m{u}\not\equiv\m{0}$ and the pre-filter~\eqref{eq:prefilter} is always active in our experiments. We use $\tau_v=0.02$\,s and $\tau_\omega=0.05$\,s. The continuous pre-filter ODE is discretized by explicit Euler at the TDS step $\Delta t=0.02$\,s, so the augmented plant actually simulated is a sampled-data approximation of~\eqref{eq:aug_DAE} within one step of the continuous construction; Corollary~\ref{cor:sampled_data_inflation} bounds the resulting safe set inflation. The new command $\m{\nu}$ is box constrained per channel to reflect realistic supervisory authority.

\vspace{0.1cm}
\noindent\textbf{Symbolic coefficient backend.} The HOCBF data $(A_r,\m{B}_r,\m{\Gamma}_r,\pi_{r-1})$ require iterated Lie derivatives of $h$ up to order $r$. Nested finite differences are unusable for $r\geq 4$ because roundoff scales as $\varepsilon/h^r$ and dominates once disturbance derivatives are nonzero. We therefore use a CasADi backend~\cite{andersson2018casadi}: ANDES residuals $\m{f},\m{g}$ are reconstructed as an MX graph, the lifted drift is built by solving the linearized algebraic constraint symbolically, and $(A_r,\m{B}_r,\m{\Gamma}_r)$ are formed by repeated \verb|ca.jtimes| applications. The resulting function is compiled with common subexpression elimination, cached to disk, and evaluated online.

\vspace{0.1cm}
\noindent\textbf{Relative degree audit.} Definition~\ref{def:rel_deg} requires uniform relative degree on an open lifted domain. Rather than assume this, we audit numerically by evaluating the symbolic lifted Lie derivative tensor at the nominal operating point and at 30 random samples in a 5\% perturbation neighborhood, reporting the smallest order at which the control direction Lie derivative stays away from zero uniformly. This is a neighborhood certificate, not a full domain certificate, and we scope the guarantees accordingly.

\vspace{0.1cm}
\noindent\textbf{Online QP and active set screening.} At each step the filter assembles the HOCBF rows, solves the soft QP~\eqref{eq:QP} with \verb|quadprog|, uses the closed form single-row soft QP update~\eqref{eq:closed_form} when one row is active and the actuator bounds are inactive, and writes the pre-filter output into ANDES. A barrier is sent to the QP only if $h_j(\m{x})\leq 0.95\,h_{j,\max}$; disabling this screening changes the minimum over buses barrier trajectory by RMS $<2\times 10^{-4}$ on the tested systems.

\vspace{0.1cm}
\noindent\textbf{Hybrid events.} The smooth theory of Section~\ref{sec:CBF} applies between switching events. For the generator trip scenarios we build a separate CasADi context on each side of the trip and swap atomically at the event. Post-switch Lie derivatives agree with tight time series reconstructions of $\dot h$ to within $3\times 10^{-8}$.

\vspace{0.1cm}
\noindent\textbf{Forward invariance in practice.} Theorem~\ref{thm:forward_inv} guarantees $h(\m{x}(t))\geq 0$ only when $\mathcal{U}_h$ is nonempty, equivalently $\rho_h\geq0$ by Theorem~\ref{thm:robust_row_feasibility}. Under tight actuator bounds and simultaneous barrier violations, $\mathcal{U}_h$ can be empty; the soft QP then uses nonzero slack, and Corollary~\ref{cor:slack_inflation} bounds the residual violation. The time domain studies include both regimes, while the reachability certificates apply to the hard safety regime required by the theorem.

\vspace{0.1cm}
\noindent\textbf{Barrier selection.} We supervise all generator speeds through the governor channel, since every generator has a local $p_{\mathrm{aux}}$ input and starts at $\omega=1$ exactly. For voltage we supervise generator terminal buses whose equilibrium voltage lies strictly inside the $[0.95,1.05]$ band; buses already outside the band at the power flow solution are excluded, because supervising a barrier initialized in its complement places the filter in the recovery regime outside the scope of Theorem~\ref{thm:forward_inv}. Buses without local generators have no local voltage actuator and cannot be supervised here. For Kundur this gives 4 voltage and 4 frequency barriers, and for IEEE-39, 7 voltage (excluding buses 35 and 36, which start below 0.95\,p.u.) and 10 frequency barriers.

\subsection{Time Domain Case Studies}\label{subsec:time_domain}

The time domain studies stress test the online filter and document its behavior when actuator authority is sufficient and when it is not. These are closed-loop ANDES simulations with the QP filter in the loop. Table~\ref{tab:safety_comparison} summarizes the safety outcome and Table~\ref{tab:runtime_performance} the online cost. Throughout, $\min h$ is the worst case over supervised elements and over the horizon for the relevant barrier family. Voltage only cases report $\min_{j,t} h_{v,j}$ with $h_{v,j}=(v_j-v_{\min})(v_{\max}-v_j)$, frequency only cases report $\min_{i,t} h_{\omega,i}$ with $h_{\omega,i}=\Delta\omega_{\max}^2-(\omega_i-\omega_0)^2$, and the Kundur combined case reports $\min(\min_{j,t} h_{v,j},\min_{i,t} h_{\omega,i})$, attained on the voltage family. Negative values denote excursions outside $\mathcal{C}$ in~\eqref{eq:safe_set}.

\input{tables/table3_safety_comparison.tex}

\begin{figure*}[t]
    \centering
    \includegraphics[width=0.9\textwidth]{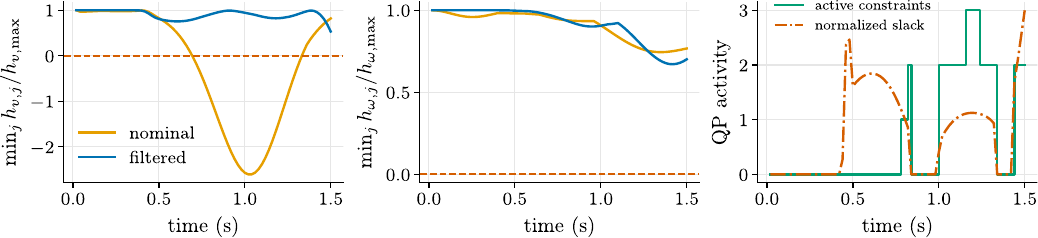}
\caption{Kundur combined load ramp case. Left: worst normalized voltage barrier, where the nominal run violates and the filtered run remains nonnegative. Middle: worst normalized frequency barrier, safe in both modes. Right: filtered QP activity, active constraints, and normalized slack.}
    \label{fig:td_kundur_combined}
\end{figure*}

\begin{figure}[t]
    \centering
    \includegraphics[width=0.9\columnwidth]{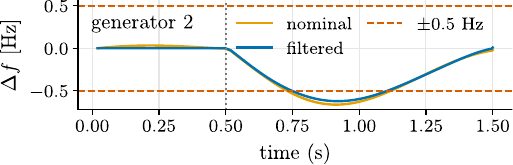}
    \caption{Kundur generator trip frequency response for the worst generator. The filtered trajectory reduces the frequency excursion relative to nominal, but both traces cross the $\pm0.5$\,Hz band, so this case is best effort rather than hard safe.}
    \label{fig:td_kundur_frequency}
\end{figure}

\vspace{0.1cm}
\noindent\textbf{Kundur.} In the voltage only load ramp the nominal trajectory stays barely inside the band, with minimum barrier $6.70\times 10^{-4}$, while the filtered run raises the minimum margin to $1.72\times 10^{-3}$. The filter shifts the supervisory references enough to improve margin without changing the qualitative response. In the generator trip frequency case the disturbance is more severe relative to governor authority. The nominal run reaches $\min h_\omega=-5.32\times 10^{-5}$ and the filtered run improves this to $-3.87\times 10^{-5}$, a reduction in peak violation rather than a hard certificate. Figure~\ref{fig:td_kundur_frequency} plots the same event in physical units. The filtered response reduces the dip, but the QP uses slack and Theorem~\ref{thm:forward_inv} does not apply because $\mathcal{U}_h$ is empty under the imposed bounds.

The combined Kundur load ramp in Figure~\ref{fig:td_kundur_combined} supervises the four voltage and four frequency barriers at once with both $v_{\mathrm{ref}}$ and $p_{\mathrm{aux}}$ channels active. The nominal trajectory violates the voltage barrier by $6.51\times 10^{-3}$, while the filtered trajectory keeps all monitored barriers nonnegative. The QP is not a trivial no-op here, since up to three constraints are active and slack is used during the most constrained part of the ramp. The case shows that the same DAE-HOCBF construction handles mixed barrier families and mixed controller channels on one augmented DAE.

\begin{figure}[t]
    \centering
    \includegraphics[width=0.9\columnwidth]{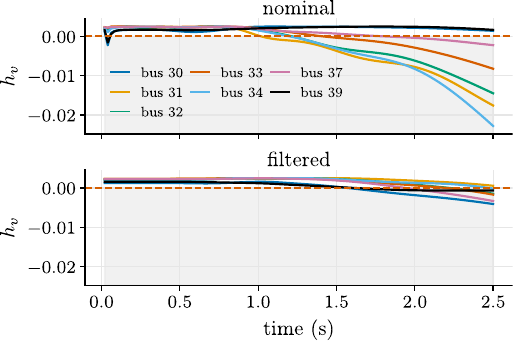}
    \caption{IEEE-39 moderate voltage load ramp. The filtered run reduces the worst voltage barrier violation from $2.28\times10^{-2}$ to $4.07\times10^{-3}$, but actuator limits still force an authority limited residual violation.}
    \label{fig:td_ieee39_voltage_moderate}
\end{figure}

\begin{figure}[t]
    \centering
    \includegraphics[width=0.9\columnwidth]{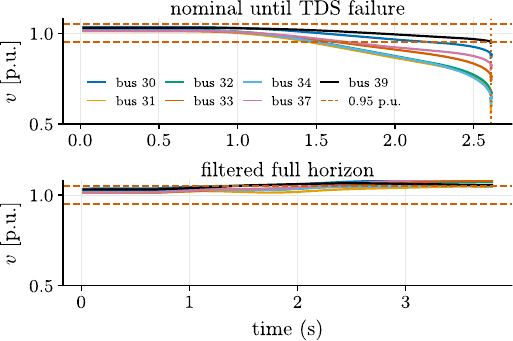}
\caption{IEEE-39 severe voltage load ramp. The nominal simulation undergoes low voltage collapse and terminates at $t=2.61$\,s, while the filtered run survives the full horizon. The filtered trajectory still enters the voltage warning band, so this is collapse prevention in the authority limited regime rather than a hard safety certificate.}
    \label{fig:td_ieee39_voltage_severe}
\end{figure}

\vspace{0.1cm}
\noindent\textbf{IEEE-39.} The IEEE-39 frequency trip is included as a deliberate no-op result: the native governor response stays inside the $\pm0.5$\,Hz band, so the filter activates zero constraints and the supervisory command stays zero. The voltage cases are more informative. For the moderate ramp, Figure~\ref{fig:td_ieee39_voltage_moderate} shows all seven supervised voltage barriers; the filter reduces the worst violation by roughly a factor of $5.6$ but cannot eliminate it under $|\nu_v|\leq0.10$. For the severe ramp, Figure~\ref{fig:td_ieee39_voltage_severe} shows the nominal ANDES simulation collapsing before the horizon, while the filtered trajectory remains numerically viable and avoids low voltage collapse. We report this as collapse prevention rather than pointwise voltage band enforcement.

\input{tables/table4_runtime_performance.tex}

The computational pattern in Table~\ref{tab:runtime_performance} matches the theory. The QP solve is not the bottleneck, at $0.27$\,ms/step even in the IEEE-39 severe voltage case; coefficient assembly for high relative degree voltage barriers is. Average callback time stays below the $20$\,ms TDS budget in every case, while peak callbacks exceed it during nonlinear portions of voltage ramps. This supports a soft real time supervisory interpretation; hard real time deployment would require command holds between coefficient refreshes or a tighter symbolic backend.

\subsection{Reachability Case Studies}\label{subsec:reach_impl}
We now turn to reachability. Section~\ref{sec:CORA} reduces verification to reachable set computation on the closed-loop augmented DAE~\eqref{eq:CL_DAE}. We implement a specialized zonotope engine, scope the admissible disturbance class, and report certified safe results on Kundur and IEEE-39.

\vspace{0.1cm}
\noindent\textbf{Engine.} The reachable set is propagated as a zonotope in the augmented state $\m{z}=[\m{x}_d^\top,\m{\eta}_0^\top,\ldots,\m{\eta}_{r_{\max}-1}^\top]^\top$. At each step we enclose the current zonotope in an interval box, evaluate $(\m{J}_a,\m{J}_d,\partial\m{g}/\partial\m{w})$ at the center via the CasADi backend, factor $\m{J}_a$ once, and eliminate algebraic increments by $\delta\m{y}=-\m{J}_a^{-1}(\m{J}_d\,\delta\m{x}_d+\partial\m{g}/\partial\m{w}\,\delta\m{\eta}_0)$. The reduced linear system in $\m{z}$ is propagated by matrix exponential integration, with additive interval remainders covering the second order Lagrange remainder of the differential flow, the algebraic resolvent linearization residual, and the closed-loop contribution of the QP filter. Girard-style order reduction caps zonotope complexity at 50 generators per state dimension. The engine reuses the same CasADi symbolic graph compiled for the online filter, so the offline build does not duplicate the symbolic work of Section~\ref{subsec:impl}.

For the filter contribution we use two modes. In \emph{inactive QP} mode the QP returns $\m{\nu}=\m{0}$ throughout the horizon, so the closed-loop flow reduces to the open-loop plant plus the native ANDES controllers, which is exact. In \emph{bounded command} mode, used when the QP may engage, we enclose $\m{\nu}(t)$ as a bounded exogenous signal in $[\m{\nu}_{\min},\m{\nu}_{\max}]$ inside the reach engine. This is sound by construction, since the QP output always respects its box, and the resulting tube is a valid outer approximation of the closed-loop reachable set.

\vspace{0.1cm}
\noindent\textbf{Why we scope the disturbance class.} Assumption~\ref{assump:main}\textit{(iii)} defines $\mathcal{W}_{r_{\max}}$ by componentwise bounds on the disturbance and its derivatives. On paper this class is the verification target, but in practice the global bounds implied by a nominal disturbance profile admit physically meaningless trajectories. For the Kundur voltage load ramp, the peak fourth derivative of the nominal smoothstep is $\bar w_4 \approx 1.95\times 10^4/\mathrm{s}^4$, and a signal holding this for 1.5\,s integrates to $|\eta_0|\approx 4\times 10^3$ in load perturbation fraction, four orders of magnitude beyond any realistic load swing. Direct Monte Carlo sampling from this class drives the power flow infeasible within one integration step. Following published reachability practice for power systems~\cite{althoff2014TPWRS}, we verify over a physically motivated subclass, a forecast uncertainty tube around the nominal trajectory,
\begin{equation}\label{eq:forecast_class}
\begin{aligned}
\mathcal{W}^{\text{fcst}}_{r_{\max}}(\varepsilon_{0:r_{\max}})
:=\{&w(t)=w_{\text{nom}}(t)+\delta(t)\;|\; \\
&|\delta^{(\ell)}(t)|\leq \varepsilon_\ell,\;\ell=0,\ldots,r_{\max}\},
\end{aligned}
\end{equation}
with $\varepsilon_\ell$ chosen so the integrated deviation stays at a specified fraction of nominal. This matches how utilities assess safety margins, namely against forecast error around a planned profile rather than against adversarial top derivative walks.

\begin{figure}[t]
    \centering
    \includegraphics[width=\columnwidth]{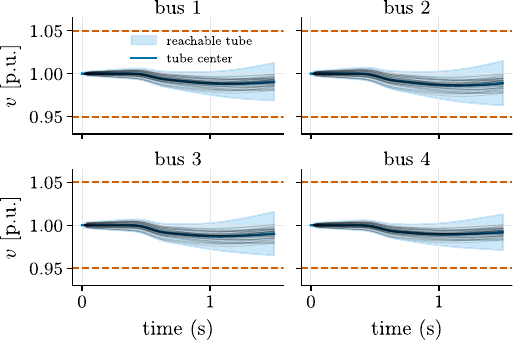}
    \caption{Reachability certificate for the benign Kundur voltage load ramp under a 1\% forecast uncertainty tube. The zonotope tube contains all 50 sampled trajectories and remains within the $[0.95,1.05]$ voltage safety band.}
    \label{fig:reach_kundur_voltage}
\end{figure}

\vspace{0.1cm}
\noindent\textbf{Certified safety cases.} We report two cases where~\eqref{eq:forecast_class} with a 1\% integrated deviation yields certified safety on the full horizon, with 50 randomly sampled trajectories all inside the computed tube. First, Kundur voltage benign (2\% load ramp over 0.4\,s, horizon 1.5\,s, all four buses supervised). The engine certifies safe with worst bus voltage margin $1.31\times 10^{-2}$\,p.u., final tube widths $[0.044, 0.052, 0.051, 0.042]$\,p.u., and 50/50 trajectories contained. Figure~\ref{fig:reach_kundur_voltage} shows the tube, the samples, and the voltage bounds.

\begin{figure}[t]
    \centering
    \includegraphics[width=\columnwidth]{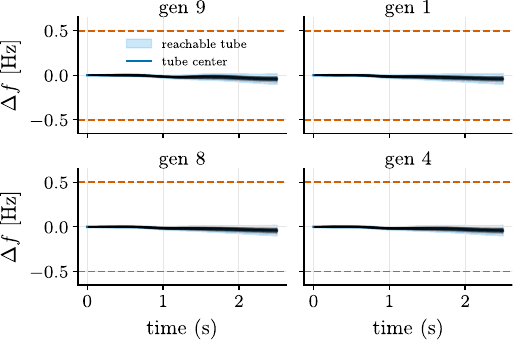}
\caption{Reachability certificate for the benign IEEE-39 frequency load ramp under a 1\% forecast uncertainty tube. The panels show the four worst margin generators; all ten speed outputs were verified, 50/50 samples were contained, and the tube remains inside the $\pm0.5$\,Hz band.}
    \label{fig:reach_ieee39_frequency}
\end{figure}

Second, IEEE-39 frequency benign (2\% load ramp over 0.8\,s, horizon 2.5\,s, all ten generator speeds supervised). The engine certifies safe with worst frequency margin $0.400$\,Hz against the $\pm0.5$\,Hz band, final tube widths $0.107$ to $0.117$\,Hz across the ten generators, and 50/50 contained. Figure~\ref{fig:reach_ieee39_frequency} shows the four worst margin generators, though all ten are checked in the certificate. This spans a system with 10 generators, algebraic dimension above 170, and relative degree 3.

The stress scenarios of Section~\ref{subsec:time_domain} operate in the authority limited regime where $\mathcal{U}_h$ is empty, the filter uses nonzero slack, and $h\geq 0$ does not hold pointwise. The hypothesis of Theorem~\ref{thm:forward_inv} fails there, so reachability cannot certify hard safety. The IEEE-39 severe voltage case still shows the filter preventing collapse, but it is not certifiable against the $\pm5\%$ band because the filtered trajectory transiently enters it. Certification and best effort behavior are complementary: the former applies when the filter has authority, while the time domain results show reduced violation and collapse prevention when it does not.

\vspace{0.1cm}
\noindent\textbf{Cost.} The reach engine is offline only. Per-step cost is dominated by the matrix exponential and the algebraic resolvent symbolic solve. On IEEE-39 frequency, average per-step wall clock is $4.0$\,s with a maximum of $6.5$\,s, and the 125-step horizon is verified in about 8 minutes. This is practical for offline certification but not for online use, where the filter of Section~\ref{subsec:impl} runs in real time.

\subsection{Limitations and Scalability}\label{subsec:limits}

We close with the main gaps between theory, implementation, and large system scaling.

\vspace{0.1cm}
\noindent\textbf{Scope.} The theory is developed on smooth segments between switching events. Generator trips are handled as two smooth segments joined by a context swap, while hard limiter saturation, anti-windup, and deadbands are not handled and the scenarios were designed to avoid them on the segments of interest. The relative degree audit is a 30-sample neighborhood certificate in a 5\% box, not a full uniform-domain proof. Reachability certification additionally requires the hard safety regime ($\mathcal{U}_h$ nonempty); in the authority limited regime the filter degrades to best effort and reachability does not apply, by design of Theorem~\ref{thm:forward_inv}.

\vspace{0.1cm}
\noindent\textbf{Online cost.} Average online coefficient assembly is $0.9$\,ms/step on Kundur voltage ($r=4$, 4 barriers) and $7.1$\,ms/step on IEEE-39 voltage ($r=5$, 7 barriers, 10 channels), well inside the $\Delta t=20$\,ms budget. Peak step times occasionally exceed the budget during the most nonlinear part of a disturbance (max $38$\,ms on IEEE-39 voltage), which is acceptable for soft real time supervisory control but for hard real time deployment requires a faster control rate or a command hold strategy at peaks. The QP solve itself is under $0.2$\,ms in all cases. We do not include an IEEE-39 combined voltage-frequency case as a main result, since it adds the expensive $r=5$ voltage barriers plus extra frequency constraints without a distinct safety story beyond the separate IEEE-39 voltage and frequency studies.

\vspace{0.1cm}
\noindent\textbf{Offline scaling wall.} The bottleneck is the one-time symbolic build of the lifted drift, which solves the algebraic constraint symbolically and whose cost grows faster than linearly in the algebraic dimension. At $n_a \approx 60$ (Kundur) the build takes seconds, at $n_a \approx 170$ (IEEE-39) it takes minutes, and preliminary tests at $n_a \approx 600$ (the 140-bus NPCC system at $r=5$) did not complete in practical time. Paths forward include compositional reachability~\cite{el2017compositional}, structured sparse symbolic solves that exploit network topology, and deploying only lower relative degree supervision on larger networks. The contribution here is the framework and its validation on benchmark DAEs; scaling to system operator size networks is left to future work.

\section{Conclusion and Future Work} \label{sec:conclusion}

We presented a unified safety filtering and verification framework for power systems modeled as smooth semi-explicit index-1 DAEs. The framework couples an online DAE-aware HOCBF QP filter, static in the supervisory input under an actuator pre-filter when needed, with offline reachability certification on the same closed-loop augmented DAE. The forward invariance theorem holds at arbitrary relative degree under absolutely continuous disturbances with an essentially bounded top derivative. For box constrained inputs we gave an exact feasibility test for each barrier row, a computable input-to-state safety margin that bounds the safe set inflation caused by slack or sampling, and a local Lipschitz guarantee for the general soft QP filter, with a closed form soft QP update in the single-row, inactive-box case. The reachability engine certifies hard safety on forecast uncertainty disturbance classes for benign scenarios on the Kundur and IEEE 39-bus benchmarks. Future work includes extending the framework to piecewise-smooth and hybrid regimes covering limiters, deadbands, anti-windup, and protection logic, applying it to inverter-based resources and grid-forming inverters, and improving scalability through faster symbolic backends, sparse algebraic resolvent solves, and compositional decomposition.

\bibliographystyle{elsarticle-num}
\bibliography{references}

\end{document}

%% file: preamble.tex
\usepackage{amsthm}
\PassOptionsToPackage{intlimits,fleqn}{amsmath}
\usepackage[T1]{fontenc}
\usepackage[utf8]{inputenc}

\usepackage{graphicx}
\usepackage{textcomp}
\usepackage{epsfig} 
\usepackage{bm}
\usepackage{amssymb}
\usepackage{url}
\usepackage{enumitem} %defined already in ieeeconf
\usepackage{multirow}
\usepackage{hhline}
\usepackage{booktabs}
\usepackage{mathtools}
\usepackage{makecell}
\usepackage{comment}

\makeatother
\usepackage{stackengine}

\newcommand{\m}{\boldsymbol}
\allowdisplaybreaks[4]
\usepackage[colorlinks = true,
linkcolor = blue,
urlcolor  = black,
citecolor = blue,
anchorcolor = blue]{hyperref}
\usepackage[framemethod=TikZ]{mdframed}
\mdfdefinestyle{MyFrame}{%
	linecolor=black,
	outerlinewidth=1.25pt,
	roundcorner=1.25pt,
	innerrightmargin=5pt,
	innerleftmargin=5pt,}

\usepackage[noabbrev]{cleveref}

\usepackage{mathtools}

\DeclarePairedDelimiter\abs{\lvert}{\rvert}%
\DeclarePairedDelimiter\norm{\lVert}{\rVert}%

\makeatletter
\let\oldabs\abs
\def\abs{\@ifstar{\oldabs}{\oldabs*}}
\let\oldnorm\norm
\def\norm{\@ifstar{\oldnorm}{\oldnorm*}}
\makeatother

\usepackage[english]{babel}
\usepackage[utf8]{inputenc}
\usepackage[super]{nth}

\usepackage{graphicx}
\usepackage{float}
\usepackage[caption = false]{subfig}

\usepackage{array}
\usepackage{threeparttable}

\usepackage[english]{babel}
\usepackage[utf8]{inputenc}
\usepackage[super]{nth}

\usepackage{lettrine} %To make first letter a two line letter and bold in the introduction

\usepackage{balance} % To balance the references in the final page

\usepackage{titlesec} % Adjust section spacing
\titlespacing*{\section}{0pt}{*0.3}{*0.2}
\titlespacing*{\subsection}{0pt}{*0.3}{*0.2}
\titlespacing*{\subsubsection}{0pt}{*0.3}{*0.2}
\AtBeginDocument{%
  \setlength{\mathindent}{0pt}%
  \setlength{\abovedisplayskip}{3.7pt}%
  \setlength{\belowdisplayskip}{3.7pt}%
  \setlength{\abovedisplayshortskip}{3.5pt}%
  \setlength{\belowdisplayshortskip}{3.5pt}%
  \setlength{\textfloatsep}{1pt plus 1pt minus 2pt}%  % between top/bottom float and body text
  \setlength{\floatsep}{1pt plus 1pt minus 1pt}%      % between consecutive floats on a float page
  \setlength{\intextsep}{1pt plus 1pt minus 1pt}%     % around h-placed (in-text) floats
  \setlength{\abovecaptionskip}{1pt}%                 % between float body and its caption
  \setlength{\belowcaptionskip}{1pt}%                 % below the caption
}

\usepackage{hyperref}
\definecolor{LCSS}{rgb}{0,0.502,0.675} %Automatica blue.
\AtBeginDocument{%
  \hypersetup{
    colorlinks  = true,
    linkcolor   = LCSS,
    citecolor   = LCSS,
    urlcolor    = LCSS,
    anchorcolor = LCSS,
    filecolor   = LCSS,
  }%
}

\usepackage{tikz}
\usetikzlibrary{shapes.geometric, arrows.meta, decorations.pathmorphing, decorations.markings, calc, 3d}
\usetikzlibrary{positioning, fit, backgrounds}
\tikzset{
	tangent/.style={decoration={
			markings,mark=at position #1 with {
				\coordinate (ta) at (0,0);
				\coordinate (tb) at (0.1,0);
			}
		},postaction=decorate},
	tangent/.default=0.5
}

\definecolor{Vgray}{RGB}{247,247,247}
\definecolor{Vedge}{RGB}{200,200,200}

\definecolor{Sub1purple}{RGB}{200,200,255}
\definecolor{Sub1edge}{RGB}{205,215,238}

\definecolor{Sub2yellow}{RGB}{251,234,206}
\definecolor{Sub2edge}{RGB}{251,234,206} % Same for fill and draw

\definecolor{Sub3green}{RGB}{215,238,210}
\definecolor{Sub3edge}{RGB}{215,232,210}

\definecolor{Sub1color}{rgb}{0.85, 0.9, 1}
\definecolor{Sub2color}{rgb}{1, 0.9, 0.7}
\definecolor{Sub3color}{rgb}{0.8, 1, 0.8}
\definecolor{Sub4color}{RGB}{200,200,255}

\definecolor{Sub4edge}{RGB}{205,215,238}

\definecolor{Sub5color}{RGB}{255,210,210}
\definecolor{Sub5edge}{RGB}{240,180,180}

\definecolor{Sub5edge}{RGB}{140,120,80}
\definecolor{Sub5color}{RGB}{255,245,225}

\definecolor{Sub6edge}{RGB}{100,130,150}
\definecolor{Sub6color}{RGB}{220,240,255}

\definecolor{Sub7edge}{RGB}{130,100,150}
\definecolor{Sub7color}{RGB}{245,230,255}

\definecolor{Sub8edge}{RGB}{150,130,100}
\definecolor{Sub8color}{RGB}{255,250,230}

\definecolor{Sub9edge}{RGB}{90,140,140}
\definecolor{Sub9color}{RGB}{220,250,250}

\definecolor{Sub10edge}{RGB}{110,110,110}
\definecolor{Sub10color}{RGB}{240,240,240}
\definecolor{LCSS}{rgb}{0,0.502,0.675} %Automatica blue.

\newcommand{\theoref}[1]{\hyperref[#1]{Theorem~\ref*{#1}}}
\newcommand{\lemref}[1]{\hyperref[#1]{Lemma~\ref*{#1}}}
\newcommand{\coref}[1]{\hyperref[#1]{Corollary~\ref*{#1}}}
\newcommand{\propref}[1]{\hyperref[#1]{Proposition~\ref*{#1}}}
\newcommand{\defref}[1]{\hyperref[#1]{Definition~\ref*{#1}}}
\newcommand{\remref}[1]{\hyperref[#1]{Remark~\ref*{#1}}}
\newcommand{\asmpref}[1]{\hyperref[#1]{Assumption~\ref*{#1}}}
\newcommand{\secref}[1]{\hyperref[#1]{Section~\ref*{#1}}}
\newcommand{\subsecref}[1]{\hyperref[#1]{Section~\ref*{#1}}}
\renewcommand{\eqref}[1]{\hyperref[#1]{(\ref*{#1})}}
\newcommand{\figref}[1]{\hyperref[#1]{Fig.~\ref*{#1}}}
\newcommand{\tabref}[1]{\hyperref[#1]{Table~\ref*{#1}}}
\newcommand{\probref}[1]{\hyperref[#1]{Problem~\ref*{#1}}}

\usepackage{caption} 
\usepackage{etoolbox}

\renewenvironment{proof}{%
	\par\vspace{1pt}%
	\noindent\textbf{\textit{\textcolor{LCSS}{Proof.}}}\enspace%
}{%
	\hfill$\blacksquare$%
	\par\vspace{1pt}%
}

\newtheorem{theorem}{\textbf{\textcolor{LCSS}{Theorem}}}
\newtheorem{mylem}{\textbf{\textcolor{LCSS}{Lemma}}}
\newtheorem{mydef}{\textbf{\textcolor{LCSS}{Definition}}}

\newtheorem{myrem}{\textbf{\textcolor{LCSS}{Remark}}}
\newtheorem{asmp}{\textbf{\textcolor{LCSS}{Assumption}}}
\newtheorem{mycor}{\textbf{\textcolor{LCSS}{Corollary}}}
\newtheorem{myprs}{\textbf{\textcolor{LCSS}{Proposition}}}

\newtheorem{problem}{\textbf{\textcolor{LCSS}{Problem}}}

\makeatletter
\def\thm@space@setup{%
  \thm@preskip=1pt plus 1pt minus 1pt%
  \thm@postskip=1pt plus 1pt minus 1pt%
}
\def\@begintheorem#1#2{%
	\item[\hskip\labelsep\textbf{\textcolor{LCSS}{\textit{#1 #2}}}]%
	\@ifnextchar[{\@withinfo}{\textbf{\textcolor{LCSS}{.}}\enspace\ignorespaces}}
	
	\def\@withinfo[#1]{% Ensure optional argument stays inline and formatted correctly
\textbf{\textcolor{LCSS}{\textit{ (#1)}}}\textcolor{LCSS}{.} \ignorespaces}
\makeatother

%% file: tables/table3_safety_comparison.tex
\begin{table*}[t]
\centering
\setlength{\tabcolsep}{3.5pt}
\renewcommand{\arraystretch}{1.15}
\caption{Safety and performance comparison for the time-domain case studies. The reported $\min h$ is the worst-case value of the safety output appropriate to each case (voltage $h_v$, frequency $h_\omega$, or combined), as defined in the text. Positive values mean the hard safety set is maintained; negative values indicate authority-limited best-effort behavior.}
\label{tab:safety_comparison}
\resizebox{\textwidth}{!}{%
\begin{tabular}{lccccccc}
\toprule
Case & min $h$ nominal & min $h$ filtered & viol. nominal & viol. filtered & max slack & max $\|\nu\|$ & Outcome \\
\midrule
Kundur voltage & $6.70\!\times\!10^{-4}$ & $1.72\!\times\!10^{-3}$ & $0$ & $0$ & $4.37\!\times\!10^{2}$ & $0.200$ & margin improved \\
Kundur frequency & $-5.32\!\times\!10^{-5}$ & $-3.87\!\times\!10^{-5}$ & $5.32\!\times\!10^{-5}$ & $3.87\!\times\!10^{-5}$ & $0.034$ & $0.057$ & best-effort \\
Kundur combined & $-6.51\!\times\!10^{-3}$ & $1.33\!\times\!10^{-3}$ & $6.51\!\times\!10^{-3}$ & $0$ & $5.34\!\times\!10^{2}$ & $0.117$ & combined safe, slack active \\
IEEE-39 frequency & $5.18\!\times\!10^{-5}$ & $6.74\!\times\!10^{-5}$ & $0$ & $0$ & $0$ & $0$ & no-op \\
IEEE-39 voltage moderate & $-0.023$ & $-4.07\!\times\!10^{-3}$ & $0.023$ & $4.07\!\times\!10^{-3}$ & $3.70\!\times\!10^{3}$ & $0.316$ & authority-limited \\
IEEE-39 voltage severe & \multicolumn{1}{c}{collapsed at $2.61$ s} & $-5.71\!\times\!10^{-3}$ & \textemdash & $5.71\!\times\!10^{-3}$ & $4.09\!\times\!10^{3}$ & $0.316$ & collapse prevented \\
\bottomrule
\end{tabular}%
}
\end{table*}

%% file: tables/table4_runtime_performance.tex
\begin{table*}[t]
\centering
\setlength{\tabcolsep}{3.5pt}
\renewcommand{\arraystretch}{1.15}
\caption{Online runtime summary for filtered case studies. Timings exclude offline symbolic prebuild and offline reachability propagation. The TDS step budget is 20 ms.}
\label{tab:runtime_performance}
\resizebox{\textwidth}{!}{%
\begin{tabular}{lccrrrrrrccc}
\toprule
Case & $n_h$ & $r$ & coeff avg & coeff max & QP avg & QP max & callback avg & callback max & active avg/max & success & budget \\
 & & & \multicolumn{6}{c}{ms/step} & & \% & \\
\midrule
Kundur voltage & 4 & $4$ & $0.89$ & $26.3$ & $0.05$ & $0.19$ & $1.00$ & $26.6$ & $0.00/0$ & $100.0$ & yes/no \\
Kundur frequency & 4 & $3$ & $0.32$ & $1.07$ & $0.05$ & $0.30$ & $0.45$ & $1.53$ & $0.29/2$ & $100.0$ & yes/yes \\
Kundur combined & 8 & $4/3$ & $1.47$ & $26.0$ & $0.05$ & $0.13$ & $1.59$ & $26.2$ & $0.67/3$ & $100.0$ & yes/no \\
IEEE-39 frequency & 10 & $3$ & $0.04$ & $0.21$ & $0.01$ & $0.10$ & $0.12$ & $0.62$ & $0.00/0$ & $100.0$ & yes/yes \\
IEEE-39 voltage moderate & 7 & $5$ & $7.05$ & $37.9$ & $0.13$ & $0.32$ & $7.29$ & $38.4$ & $1.56/6$ & $100.0$ & yes/no \\
IEEE-39 voltage severe & 7 & $5$ & $14.9$ & $85.7$ & $0.27$ & $2.08$ & $15.4$ & $86.9$ & $2.48/6$ & $100.0$ & yes/no \\
\bottomrule
\end{tabular}%
}
\end{table*}